\pdfoutput=1
\documentclass[10pt,conference]{IEEEtran}
\usepackage{cite}
\usepackage{amsmath,amssymb,amsfonts}
\usepackage{algorithmic}
\usepackage{graphicx}
\usepackage{textcomp}
\usepackage{xcolor}
\usepackage[hyphens]{url}
\usepackage{fancyhdr}

\usepackage{comment}
\usepackage{algorithm}
\usepackage{mhchem}
\usepackage{tikz}
\usepackage[hidelinks]{hyperref}
\usepackage{balance}

\newcommand{\hpcayear}{2024}

\newcommand{\textfw}[1]{\scalebox{.8}[1.0]{\texttt{#1}}}

\newcommand*\circled[1]{\tikz[baseline=(char.base)]{
\node[shape=circle,fill,inner sep=0.5pt] (char) {\footnotesize \textcolor{white}{#1}};}}

\title{Bandwidth-Effective DRAM Cache for \\ GPUs with Storage-Class Memory} 

\def\hpcacameraready{} %
\newcommand\hpcaauthors{Jeongmin Hong$\dagger$, Sungjun Cho$\dagger$,  Geonwoo Park$\dagger$, Wonhyuk Yang$\dagger$, Young-Ho Gong$\star$, and  Gwangsun Kim$\dagger$\\}
\newcommand\hpcaaffiliation{\hspace{28mm}POSTECH$\dagger$\hspace{59mm}Soongsil University$\star$\\Department of Computer Science and Engineering\hspace{32mm}School of Software}
\newcommand\hpcaemail{\hspace{-15mm}\{jmhhh, allencho1222, geonwoo1998, wonhyuk,  g.kim\}@postech.ac.kr\hspace{17mm}yhgong@ssu.ac.kr}

\author{
  \ifdefined\hpcacameraready
    \IEEEauthorblockN{\hpcaauthors{}}
      \IEEEauthorblockA{
        \hpcaaffiliation{} \\
        \hpcaemail{}
      }
  \else
    \IEEEauthorblockN{\normalsize{HPCA \hpcayear{} Submission
      \textbf{\#\hpcasubmissionnumber{}}} \\
      \IEEEauthorblockA{
        Confidential Draft \\
        Do NOT Distribute!!
      }
    }
  \fi 
}

\fancypagestyle{camerareadyfirstpage}{%
  \fancyhead{}
  
  \fancyhead[C]{
    \ifdefined\aeopen
    \parbox[][12mm][t]{13.5cm}{\hpcayear{} IEEE International Symposium on High-Performance Computer Architecture (HPCA)}    
    \else
      \ifdefined\aereviewed
      \parbox[][12mm][t]{13.5cm}{\hpcayear{} IEEE International Symposium on High-Performance Computer Architecture (HPCA)}
      \else
      \ifdefined\aereproduced
      \parbox[][12mm][t]{13.5cm}{\hpcayear{} IEEE International Symposium on High-Performance Computer Architecture (HPCA)}
      \else
      \parbox[][0mm][t]{13.5cm}{\hpcayear{} IEEE International Symposium on High-Performance Computer Architecture (HPCA)}
    \fi 
    \fi 
    \fi 
    \ifdefined\aeopen 
      \includegraphics[width=12mm,height=12mm]{ae-badges/open-research-objects.pdf}
    \fi 
    \ifdefined\aereviewed
      \includegraphics[width=12mm,height=12mm]{ae-badges/research-objects-reviewed.pdf}
    \fi 
    \ifdefined\aereproduced
      \includegraphics[width=12mm,height=12mm]{ae-badges/results-reproduced.pdf}
    \fi
  }
  \fancyfoot[C]{}
}
\begin{document}
\maketitle

\ifdefined\hpcacameraready 
  \thispagestyle{camerareadyfirstpage}
  \pagestyle{empty}
\else
  \thispagestyle{plain}
  \pagestyle{plain}
\fi

\newcommand{\hpcaheight}{0mm}
\ifdefined\eaopen
\renewcommand{\hpcaheight}{12mm}
\fi

\begin{abstract}

We propose overcoming the memory capacity limitation of GPUs with
high-capacity Storage-Class Memory (SCM) and DRAM cache. 
By significantly increasing the memory capacity with SCM, 
the GPU can capture a larger fraction of the memory footprint than HBM 
for workloads that mandate memory oversubscription, resulting in substantial speedups.
However, the DRAM cache needs to be carefully designed to address
the latency and bandwidth limitations of the SCM while minimizing
cost overhead and considering GPU's characteristics.
Because the massive number of GPU threads can easily thrash the DRAM cache
and degrade performance, 
we first propose an \emph{SCM-aware DRAM cache bypass policy for GPUs} that considers the multi-dimensional characteristics of memory accesses by GPUs with SCM to bypass DRAM for data with low performance utility.
In addition, to reduce DRAM cache probe traffic and increase effective DRAM BW
with minimal cost overhead,
we propose a \emph{Configurable Tag Cache (CTC)} that repurposes part of 
the L2 cache to cache DRAM cacheline tags.
The L2 capacity used for the CTC can be adjusted by users for adaptability.
Furthermore, to minimize DRAM cache probe traffic from CTC misses,
our \emph{Aggregated Metadata-In-Last-column (AMIL)} DRAM cache organization co-locates all DRAM cacheline tags in a single column within a row.
The AMIL also retains the full ECC protection, unlike prior DRAM cache
implementation with Tag-And-Data (TAD) organization. 
Additionally, we propose SCM throttling to curtail power consumption and
exploiting SCM's SLC/MLC modes to adapt to workload's memory footprint.
While our techniques can be used for different DRAM and SCM devices,
we focus on a \emph{Heterogeneous Memory Stack (HMS)} organization
that stacks SCM dies on top of DRAM dies for high performance.
Compared to HBM, the HMS
improves performance by up to 12.5$\times$ (2.9$\times$ overall) and reduces energy by up to 89.3\% (48.1\% overall). 
Compared to prior works, we reduce DRAM cache probe and SCM write traffic 
by 91-93\% and 57-75\%, respectively.

\end{abstract}

\section{Introduction}
\label{sec:introduction}

\begin{figure}
\centering
\includegraphics[width=0.47\columnwidth]{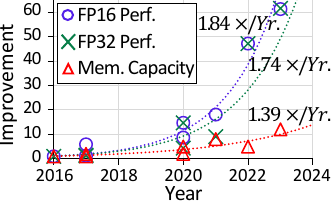} 
\hspace{.1in}
\includegraphics[width=0.45\columnwidth]{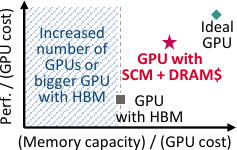}
\hbox{\hspace{0.15in}\footnotesize{ (a) \hspace{1.55in} (b)\hspace{0.01in}}}
\caption{(a) Improvement in compute throughput (with Tensor Core~\cite{tensor_core} and Matrix Core~\cite{matrix_core} 
when applicable) and memory capacity of GPUs over time~\cite{h100,a100,v100,p100,mi25,mi100,mi250x,mi300x}. 
(b) Cost-effectiveness of different GPU architectures under memory oversubscription.}
\label{fig:gpu_memory}
\end{figure}

Rapidly-increasing data size in various domains~\cite{src_memory, ai_memory_wall} %
has created huge challenges for the memory system of GPUs.
Although High-Bandwidth Memory (HBM) has been adopted to meet the high memory bandwidth (BW) requirements of GPUs, it fails to fulfill the memory capacity needs of critical workloads, such as deep learning and large-scale graph analytics.
Moreover, the memory capacity of GPUs has grown much slower than the compute throughput (Fig.~\ref{fig:gpu_memory}a).

When data size exceeds GPU memory capacity, the data must be migrated repeatedly between the CPU and GPU, either manually or automatically.
However, manual migration can be laborious for programmers, and it is infeasible for irregular workloads because the data access pattern is unpredictable. 
On the other hand, demand paging approaches 
(e.g., NVIDIA Unified Memory~\cite{um})
can automatically manage data movement, but it can significantly degrade performance due to high page fault-handling latency and limited PCIe BW~\cite{uvmsmart, um_icpe, uvmbench}. 
This overhead can be particularly severe for irregular workloads since prefetch/eviction policies become ineffective~\cite{uvm_analysis}. 
For example, the runtime of \textfw{bfs} can increase by $\sim$4.5$\times$ with only 125\% oversubscription
(i.e., exceeding memory capacity by 25\%) compared to when the GPU is not oversubscribed~\cite{adaptive}.

To avoid oversubscription, multiple GPUs can be used or bigger GPUs with more memory devices can be created.
However, they superlinearly increase system cost due to high-speed link
interface/switches~\cite{nvswitch} required and/or
sublinear pin BW scaling with area~\cite{loh_memsys}.
Thus, these approaches lower memory capacity per GPU cost compared to the baseline GPU with oversubscribed HBM (shaded region in Fig.~\ref{fig:gpu_memory}b).
Using multiple GPUs can also require significant programmer efforts (\S\ref{sec:multi-gpu}).

Meanwhile, emerging Storage-Class Memory (SCM) offers a potential solution to the capacity limitations of DRAM with its higher memory density. Recent improvements in the device characteristics of the SCM, in terms of endurance~\cite{pcm_endurance_iedm16, pcm_endurance_vlsi19}, reliability~\cite{hynix_som}, performance~\cite{hynix_som}, 
and capacity~\cite{hynix_pcm_256gb}, have also made it an even more attractive choice.
In addition, albeit slower than DRAM, SCM access is still much less expensive than accessing host memory through PCIe. 
SCM is also known to have a lower per-bit dollar cost than DRAM~\cite{optane_cost}.

However, entirely replacing GPU's DRAM with SCM would be inefficient due to the lower performance
and higher energy consumption of SCM~\cite{scm_guideline, lee2009architecting}.
Thus, DRAM has to be used together to mitigate the disadvantages of SCM.
In particular, HW-managed DRAM cache is suitable for GPUs as SW-managed scheme would incur high overhead from GPU page table update by the host-side driver~\cite{uvmsmart}. 

To this end, we propose a novel DRAM cache design for GPUs with SCM. 
By significantly increasing the memory capacity with SCM,
the GPU can avoid memory oversubscription entirely or capture a larger fraction of the memory footprint.
At the same time, the performance impact of SCM is mitigated with an effective DRAM cache design.
As a result, higher performance and memory capacity per cost can be achieved to approach  
the ideal GPU (Fig.~\ref{fig:gpu_memory}b).

We design a \emph{bandwidth-effective} DRAM cache optimized 
to meet GPU's high BW demands by minimizing the BW overhead from
DRAM caches. 
In particular, a large number of concurrent memory accesses from 100,000s of threads can easily thrash the DRAM cache and waste BW for cache fills and write-backs.
Prior work~\cite{bear, rrip_dram, redcache} on DRAM cache for CPUs proposed bypassing based on random sampling or access frequency for higher DRAM cache hit rate and lower latency. 
However, our DRAM cache requires a different bypass mechanism that considers GPU workload's access patterns (e.g., inter-thread spatial locality)
and SCM properties, especially its long write latency and high write energy~\cite{scm_guideline, lee2009architecting}.
In addition, simply maximizing DRAM cache hit rates may not enhance performance due to reduced parallelism 
across DRAM and SCM. 
Thus, we propose an \emph{SCM-aware DRAM cache bypass policy for GPUs} that captures the multi-dimensional characteristics of SCM and GPU workload's access patterns (i.e., spatial locality, read/write access type, and access frequencies of pages) for effective caching.

Despite the bypass, excessive DRAM cache probe traffic can still result from tag accesses that contend with data accesses.
To reduce the BW overhead, caching DRAM cache tags on-chip can be considered. 
However, blindly provisioning large amounts of SRAM to filter out tag accesses for large DRAM caches 
increases GPU cost without benefiting workloads that do not use the DRAM cache well (e.g., due to bypassing).

Thus, we propose a \emph{Configurable Tag Cache (CTC)} to enable adjustment of SRAM capacity used for caching DRAM cache tags. The CTC repurposes some of the L2 cache ways
to store DRAM cache tags. The user can configure the number of ways used for L2 cache and CTC, similar to configuration of L1 data cache and shared memory~\cite{a100}. 
The CTC incurs low overhead by exploiting the existing L2 cache's data array.

However, when a CTC miss occurs, multiple tags of DRAM cachelines in a row need to be fetched.
Thus, to minimize the DRAM BW overhead for tag accesses, 
we propose an \emph{Aggregated Metadata-In-Last-column (AMIL)} organization 
that co-locates all tags from a DRAM row in the last column's data portion. 
The last column is used because it tends to be underutilized when data placement 
is done in an aligned manner. 
Although the SCM data that maps to the last column has to always bypass the DRAM cache, it accounts
for a very small fraction of data (e.g., only 1.56\% of a 2048~KiB row with 32~B column of HBM) and
incurs only 1.7\% performance loss according to our study. 
The AMIL also retains the full ECC protection in the DRAM cache 
in contrast to prior work~\cite{knights_landing, alloycache, bear, carve} that has to repurpose ECC 
bits to store tags.

Among different approaches to combine SCM and DRAM for a GPU, 
we focus on the study of a \emph{Heterogeneous Memory Stack (HMS)}, which 
integrates SCM and DRAM in a 3D-stacked memory using Through-Silicon Vias (TSV).
As SCM and DRAM share the same bus in this design, the bus BW can be
flexibly utilized across varying DRAM cache hit rates (\S\ref{sec:design_space}).
However, our DRAM cache is also effective even if SCM is integrated as 
separate devices or external SCM attached with high-speed links~\cite{nvlink, cxl}.

We additionally propose power management and performance optimization
to address SCM's device characteristics. 
When memory power consumption is high, the SCM can be throttled to 
reduce power by adjusting the timing parameters. 
Consequently, the HMS power can remain below the maximum power of 
an ideal high-capacity HBM, while still outperforming an oversubscribed HBM.
In addition, when workload's footprint 
is small, the DRAM can be used as part of memory rather than a DRAM cache,
to hold the majority of the data. The remaining data can be held in the SCM
that operates in the performance-oriented SLC mode instead of capacity-oriented MLC mode.
As a result, the performance of HMS can approach that of HBM for small memory footprint.

We demonstrate the effectiveness of the HMS using various GPU workloads
that include multi-GPU large language model (LLM) training.

To summarize, we make the following contributions: 

\begin{itemize}
    \item {\bf To the best of our knowledge, this work is the first to explore the design space of the DRAM cache for GPUs with SCM.}
    Our proposed GPU memory system can overcome the limited memory capacity and
    resulting performance degradation from oversubscription of DRAM-only GPUs.
    
   \item Our \emph{Aggregated Metadata-In-Last-column (AMIL)} DRAM cache organization 
   minimizes tag probe overhead by keeping all tags in a single row 
   without compromising ECC protection as in prior works.
   
   \item We propose \emph{SCM-aware DRAM cache bypass policy for GPUs} to minimize the performance penalty of SCM by considering the memory access patterns of GPUs and the device characteristics of DRAM and SCM. 
   
   \item Our \emph{Configurable Tag Cache (CTC)} repurposes a user-specified
   portion of the L2 cache ways to store DRAM tags, substantially reducing
   DRAM tag probe overhead.   

    \item We show that our DRAM cache can significantly improve performance by up to 12.5$\times$ (2.9$\times$ overall) and reduce energy consumption by up to 89.3\% (48.1\% overall) compared to HBM, with low hardware overhead.

    \item We propose simple techniques to mitigate SCM's
    power consumption and performance impact 
    by adjusting the operation modes of the SCM and DRAM.
    
\end{itemize}

\section{Background and Motivation}

\subsection{Unified Memory}
\label{sec:um}
\begin{figure}
\centering
\includegraphics[width=0.2\linewidth]{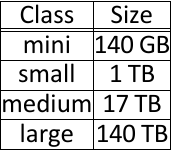}
\includegraphics[width=0.76\linewidth]{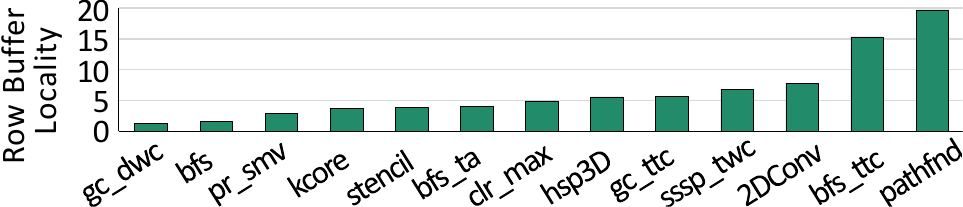} \\
\raggedright{\hspace{0.35in}\footnotesize{(a)} \hspace{1.6in} \footnotesize{(b)}}
\caption{(a) Graph500 benchmark's data size example~\cite{graph500}. (b) Row buffer locality (defined as the average number of column accesses per row activation) of representative workloads.}
\label{fig:workload_char}
\end{figure}

Modern GPUs support Unified Memory (UM)~\cite{um} that provides 
a single virtual address space for the host and device and 
automates data transfers between them without explicit copies.
UM also enables GPU memory oversubscription, allowing kernel's memory footprint to exceed GPU memory capacity.
It is especially useful for large irregular workloads under oversubscription (Fig.~\ref{fig:workload_char}a), as manual 
memory copy is infeasible for unpredictable access patterns.

When a GPU accesses data in the host memory, a page fault occurs to 
initiate page transfers or swaps between the host and device, in 4~KiB page 
granularity on x86~\cite{um_analysis}.
This process involves CUDA runtime and GPU driver on the host, and the data transfer goes through PCIe with limited BW, leading to low performance~\cite{um_analysis, um_icpe}.
To recover performance, prior works~\cite{uvmsmart, batchaware, uvmframework} proposed 
prefetch, eviction, and data transfer schemes.
For example, Tree-Based Neighborhood (TBN) prefetch and pre-eviction policies of NVIDIA GPUs adaptively migrate data in larger granularity of up to 1~MiB for high PCIe BW utilization~\cite{uvmsmart}.
The vDNN~\cite{vdnn} exploits the access patterns of activations known a priori 
for prefetching and eviction in DNN training.
However, their effectiveness can be limited for irregular workloads due to unpredictable access patterns.
Some recent GPUs~\cite{gh100} support host connectivity through NVLink with a high BW of 900~GB/s, but
oversubscription still hurts performance as we show in \S\ref{sec:perf}.

\subsection{Modeling Unified Memory}
\label{sec:um_model}

\begin{figure}
\centering
\includegraphics[width=0.92\linewidth]{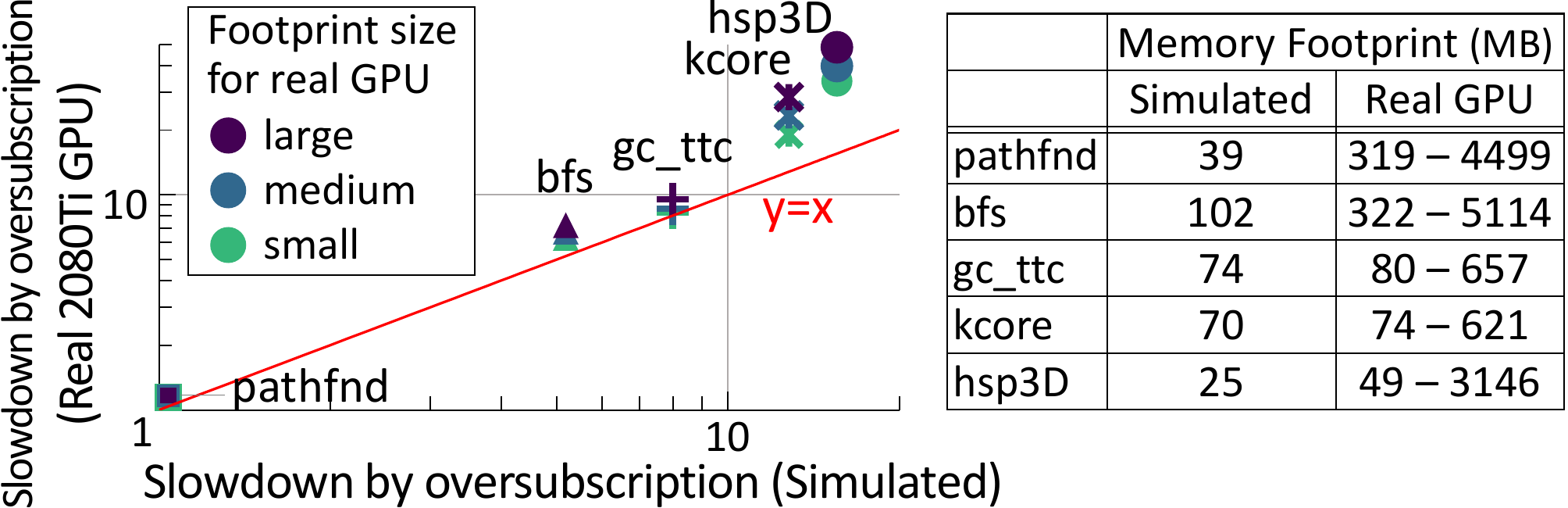}
\caption{(Left) Validation of UM simulation at a fixed oversubscription ratio (log scale plot).
(Right) Workloads' memory footprints used for validation.}
\label{fig:real_gpu_result}
\end{figure}

As computer architecture research is often done using simulators,
prior work on HW-assisted UM~\cite{uvmsmart, uvmsmart_github} modeled UM by modifying GPGPU-sim~\cite{gpgpu-sim}.
However, due to the slowdown from page faults by the GPU, the simulation speed is also 
slowed down significantly (up to 5$\times$ in our evaluation) by oversubscription.
Based on our estimation,
\textbf{simulating a full A100 GPU (80~GiB) that is oversubscribed to hold 75\% of
the memory footprint would take up to 57 years.}
Consequently, to our knowledge, all prior work on UM used
scaled-down configurations for simulation, using footprints between 15-74~MiB on average~\cite{uvmsmart, gpudmm, 2016:hpca:zheng, uvmframework, batchaware}. %
To validate the methodology, we analyzed the impact of oversubscription
on a real NVIDIA RTX 2080 Ti GPU and a simulated GPU~\cite{uvmsmart_github} for representative workloads.
For the real GPU, we induced oversubscription by 
pinning dummy data on the GPU, thereby limiting available memory to 75\% of the workload's memory footprint.
Input data were generated using~\cite{rodinia, OSMnx}. 
Results in Fig.~\ref{fig:real_gpu_result} show that real GPU exhibited similar 
or even higher slowdown from oversubscription than in simulation. 
This discrepancy can be attributed to the simulator's optimistic page-fault handling latency of 20$\mu$s, which is known to be a lower 
bound~\cite{batchaware, 2016:hpca:zheng}. 
Also, larger footprints under the same oversubscription ratio further 
slowed down the real GPU.
The simulation results for oversubscribed GPU are also consistent with
measurements on real GPU~\cite{um_vldb, um_icpe}.
Although not shown here due to space constraints, 
compute-bound workloads
tested -- 2mm~\cite{polybench} and lavamd~\cite{rodinia} -- also 
showed the same behavior, with little slowdowns.
Thus, we adopt the simulator to model UM.

\subsection{Challenges of Multi-GPU Programming}
\label{sec:multi-gpu}
If a workload's data do not fit in a GPU, multiple GPUs can be used to partition the data.
However, currently CUDA or OpenCL cannot automatically scale a single-GPU
workload to multiple GPUs. Thus, in general, the programmer has to manually modify the code to split the data and computation even for regular workloads. 
Moreover, additional kernels often have to be created to process data shared between GPUs
and communication has to be manually optimized for best performance~\cite{multi-gpu-programming}.
For irregular (e.g., graph) workloads, the code often has to be entirely rewritten
using frameworks for the target domain (e.g., WholeGraph~\cite{wholegraph} for GNN
and Pangolin~\cite{pangolin} for graph pattern mining). 
Due to the overhead of graph pre-processing, load-imbalance, and inter-GPU communication,
the performance often scales poorly with GPU count and may even be degraded~\cite{graph_study, gap, gunrock}.
Therefore, reducing the number of GPUs for large-scale workloads by increasing GPU memory capacity
can often not only reduce system cost but also improve performance. 
In addition, higher GPU memory capacity widens the range of workloads that a single GPU can execute.

\subsection{SCM Characteristics}
\label{sec:dram_cache_motivation}

SCM refers to a set of non-volatile memory (e.g., Phase Change Memory or PCM) located between DRAM and flash devices in the memory hierarchy in terms of latency, BW, and density. PCM uses phase-change material that switches between a high-resistance amorphous state (logical ``0") and a low-resistance crystalline state (logical ``1")~\cite{lee2009architecting} and is mature enough to be commercialized~\cite{emerging_mem, 3dxpoint_intel}.
For state transition, the cell is heated up to crystallization (melting) point for a SET (RESET) operation. %
PCM can also provide Multi-Level Cell (MLC) capability~\cite{lee2009architecting} and multiple decks in a die for higher capacity~\cite{3dxpoint_intel, hynix_pcm_256gb}.
Moreover, PCM can realize high 10-year data retention temperature~\cite{pcm_retention}.
Recent SCM also uses less power than conventional
3D XPoint memory~\cite{hynix_som} and several works 
showed high endurance of $10^{11}-10^{12}$ programming cycles~\cite{pcm_endurance_iedm16, pcm_endurance_vlsi19}.

Although SCM has longer row activation latency, 
the column access latency (i.e., $t_{CL}$) is the same as that of DRAM since
the row buffer access mechanism is orthogonal to memory technology~\cite{scm_guideline, lee2009architecting}. Thus, when the row buffer locality is high, even slow
memory devices can saturate the memory channel.
In addition, current DRAM devices, such as HBM and GDDR, have channel BW significantly lower than the internal BW from multiple banks in a die.
Thus, even if each bank's BW is lowered by replacing
the DRAM arrays with SCM arrays, the channel BW can still be saturated 
with high row buffer locality.

\begin{figure}
\centering
\includegraphics[width=0.95\linewidth]{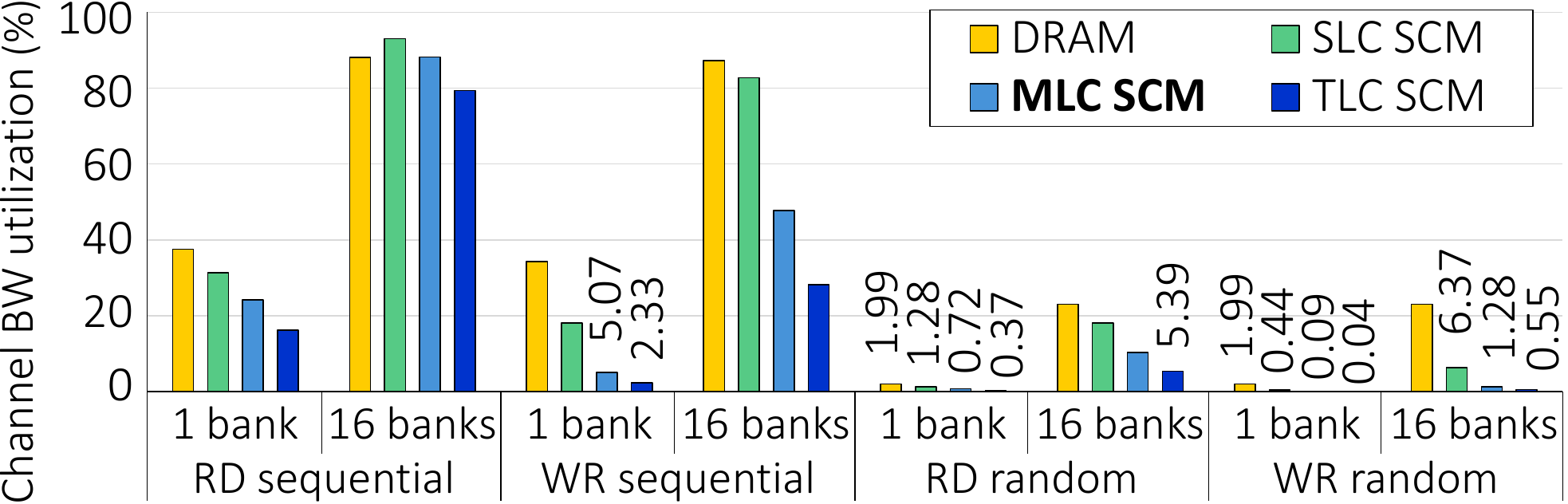}
\caption{Memory channel BW utilization from memory devices with
HBM organization for synthetic access patterns (configuration details in \S \ref{sec:methodology}).}
\label{fig:synthetic_traffic}
\end{figure}

Synthetic traffic results in Fig.~\ref{fig:synthetic_traffic} show that, for sequential read accesses over 16 banks, different SCM devices can achieve similar channel BW 
utilization as DRAM, even though single-bank BW of SCM is considerably 
lower than that of DRAM. The SLC SCM even achieves slightly higher BW than 
DRAM by eliminating refresh operations. 
Recent work~\cite{hynix_pcm_256gb} also demonstrated that SCM  
can provide a high capacity of 256~Gib (cf. 24~Gib 
DDR5 DRAM chip~\cite{ddr5_24gb}) while providing high 15~GB/s BW 
from a single chip, although its interface was not disclosed 
(cf. 51.2~GB/s peak BW from 8 chips in a consumer DDR5-6400 DIMM).
However, for streaming writes, high SCM write latencies 
result in lower overall BW even with 16 banks. Furthermore, 
for random accesses, SCM BW reduces further
due to very low locality.

While Optane DIMM with PCM exhibits low BW even for streaming accesses~\cite{optane_basic}, it can result from its multiple levels of SRAM and 
DRAM buffers within the DIMM, internal address translation, and intra-PCM data migration that can severely degrade PCM performance~\cite{PCMCsim, optane_micro20}, rather than the raw performance of PCM.
Using PCMCSim~\cite{PCMCsim} and synthetic streaming access pattern, we confirmed that, without such overhead, the PCM chip's 2$\times$DDR4-2666 interface BW can be saturated. 
LENS~\cite{optane_micro20} also reported a consistent result 
(4KiB data access from PCM in 100ns, achieving  $\sim$40 GB/s BW).

\subsection{Considerations in DRAM Cache Design for GPU with SCM}
\label{sec:considerations}

While GPUs require high memory BW, SCM throughput varies 
substantially based on access type (i.e., read or write) and locality (Fig.~\ref{fig:synthetic_traffic}).
GPU workloads also have varying access locality (Fig.~\ref{fig:workload_char}b).
Thus, characteristics of GPU workloads and SCM should be carefully 
considered for DRAM cache to hold hot data 
with low spatial locality, while cold data with high spatial locality 
resides in SCM. 
By filtering writes to SCM, DRAM cache can also mitigate 
the high write latency and energy~\cite{scm_guideline}.
However, choosing which data to cache in DRAM is challenging because of multi-dimensional access characteristics (i.e., spatial locality, hotness, and write intensity).

Most prior work on DRAM cache targeted CPUs and focused on 
minimizing latency~\cite{alloycache, unisoncache, tictoc, banshee, bear, mostly_clean}, so they can be suboptimal for GPUs, which are more sensitive to memory BW than latency~\cite{gpu_noc}. 
In addition, prior works on bandwidth-efficient DRAM cache assumed on-package DRAM cache backed by off-package DRAM~\cite{bear, tictoc, banshee}, rather than DRAM cache backed by
SCM that we assume.
Furthermore, DRAM caches are often managed in page granularity~\cite{banshee, chop, tdc, thermostat,pcm_isca09,nimble_page, hetero_os, optane_micro20, zhao2012optimizing}, but
GPUs can suffer from the resulting waste of BW. 
Also, the spatial locality that GPU exhibits \emph{across threads} in a warp or thread block 
is different from the \emph{intra-thread} spatial correlation in CPU workloads from complex 
data structures or control-flow~\cite{sms}. Thus, footprint caching~\cite{footprint_cache, ftdc, unisoncache} proposed for CPUs 
can be ineffective for GPUs. 
We discuss prior work further in \S \ref{sec:related_work}.

To our knowledge, no prior work incorporated spatial locality across GPU threads in designing the DRAM cache as we do in determining the DRAM cache design, bypass policy, and on-chip DRAM cache metadata caching mechanism.

\subsection{Feasibility of TSV-based 3D Stack of SCM and DRAM}

In TSV-based 3D integration, each die is fabricated separately, 
using different processes if needed. It avoids the manufacturing difficulties of sequentially fabricating a top die directly on a 
bottom die in monolithic 
3D~\cite{m3d_isca}. 3D-stacking of heterogeneous dies with TSV has been 
extensively studied and demonstrated for PCM~\cite{tsv}, DRAM~\cite{hbm}, CMOS 
sensors~\cite{tsv_sony}, flash devices~\cite{3D_integration_sk}, and MEMS~\cite{tsv_mems}. 
Here, we examine key considerations regarding the feasibility of HMS.

In general, TSVs can pose signal integrity issues due to coupling between a TSV and 
nearby TSVs or circuitry, as well as reliability issues arising from the high-temperature 
manufacturing process~\cite{tsv}. 
The mass production of HBMs since 2015~\cite{amd_fury_hotchips2015} demonstrates that 
these challenges have been well understood and overcome for 3D-stacked DRAM. Because the
HMS places TSVs in the same peripheral IO circuitry region 
as in HBM (Fig.~\ref{fig:design_space}a)~\cite{hbm_hynix_2017}, 
apart from the memory cell array, we do not introduce any new challenges in these 
aspects compared to HBM. In addition, SCM media, such as typical PCM with 
ovonic threshold switch (OTS)~\cite{ots-pcm, hynix_pcm_256gb}, are compatible with back-end-of-line (BEOL)
process and can withstand high-temperature TSV fabrication process~\cite{tsv_temperature}.

Additionally, power delivery 
network (PDN) with TSVs~\cite{tsv} should provide sufficient power for SCM.
Considering $\sim$10x difference in access \emph{energy} between DRAM 
($\sim$1 pJ/bit~\cite{subchannel, fine-grained-dram}) and 
PCM ($\sim$10 pJ/bit~\cite{pcm_characteristics, lee2009architecting}) and 
that PCM accesses are 10-100x slower than DRAM~\cite{pcm_characteristics, lee2009architecting, scm_guideline},  
PCMs can consume similar or less power than DRAM
per bank (i.e., power=energy/delay)~\cite{lee2009architecting}.
However, compared to DRAM, multiple SCM row accesses 
from more banks can overlap due to its longer delays,
using more power per channel. Recent DRAM with processing-in-memory capability has shown that 4-5x higher power can be supplied within HBM~\cite{samsung_hbm_pim} and that $t_{FAW}$ constraint can be removed~\cite{hynix_gddr_aim}. 
Thus, the PDN issue can be addressed similarly for SCM.

In addition, heat dissipation from SCMs in an HMS can be an 
issue for temperature-sensitive DRAM~\cite{dram_temperature}.
It is a fundamental challenge in 3D stacks, 
including HBMs, and HBMs can be throttled by the 
memory controller at high temperatures~\cite{hbm_throttle}. 
Similarly, we show that a simple SCM throttling technique can effectively 
mitigate the thermal issue (\S \ref{sec:opt}), and even without throttling, 
the worst-case peak HMS temperature differs from that of HBM by 
less than 0.1\% (\S \ref{sec:thermal}) as SCMs are 
placed in the upper rank of the stack, close to the heat sink. Recent cooling 
solutions (e.g., liquid immersion cooling adopted in production datacenters~\cite{immersion_cooling1, immersion_cooling2}) are also proven 
to be more energy-efficient while allowing processors to operate at higher power
in comparison to air cooling. Thus, they can allow 
for more aggressive SCM devices. The energy and heat issues of SCM can
also be mitigated by device scaling because the energy 
of SCMs, such as PCM, decreases with the cell material volume~\cite{pcm_scaling, pcm_scaling2}.

\section{DRAM Cache for GPUs with SCM}

\subsection{Design Space of Heterogeneous Memory}
\label{sec:design_space}

\begin{figure}
\centering
\includegraphics[width=1.0\columnwidth]{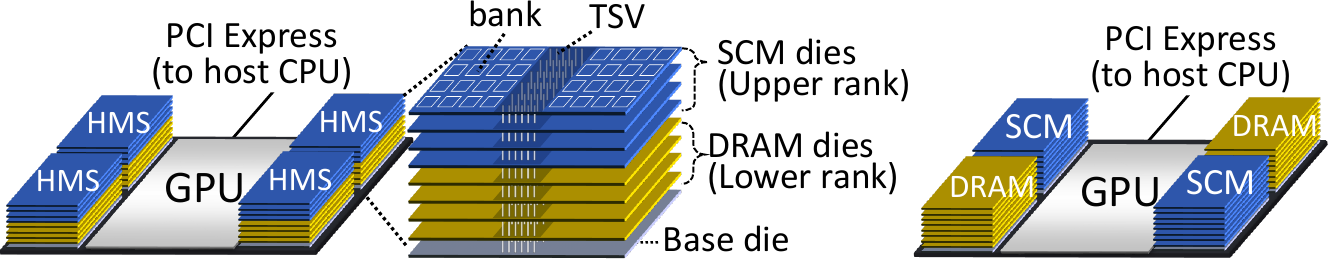}
\hbox{\hspace{0.3in}\footnotesize{(a)} \hspace{1.6in} \footnotesize{(b)}}
\caption{Design space of a GPU with SCM and DRAM cache with (a) 3D-stacked DRAM and SCM
and (b) separate DRAM and SCM stacks.}
\label{fig:design_space}
\end{figure}

\begin{figure}
\centering
\includegraphics[width=0.325\linewidth]{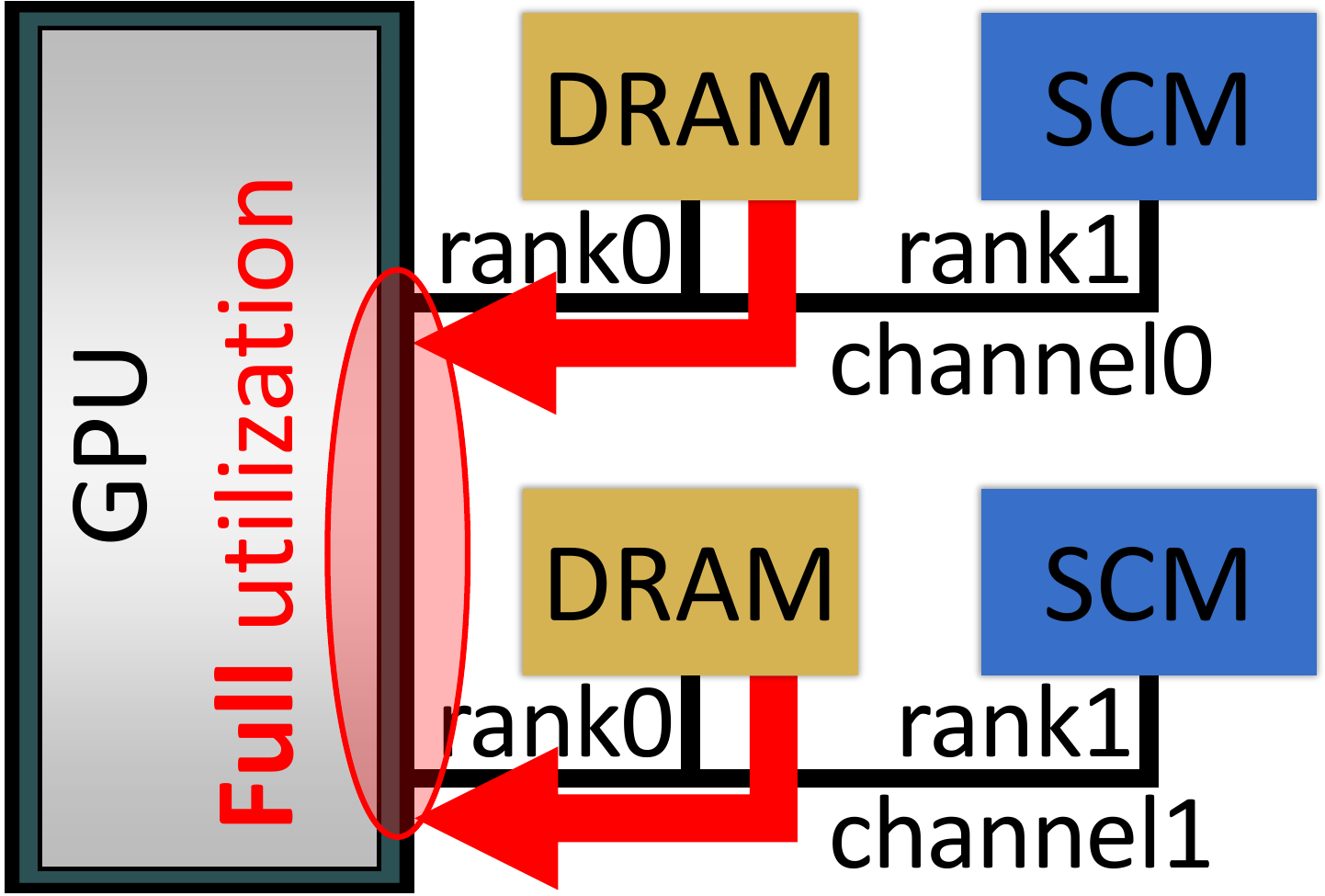} 
\includegraphics[width=0.325\linewidth]{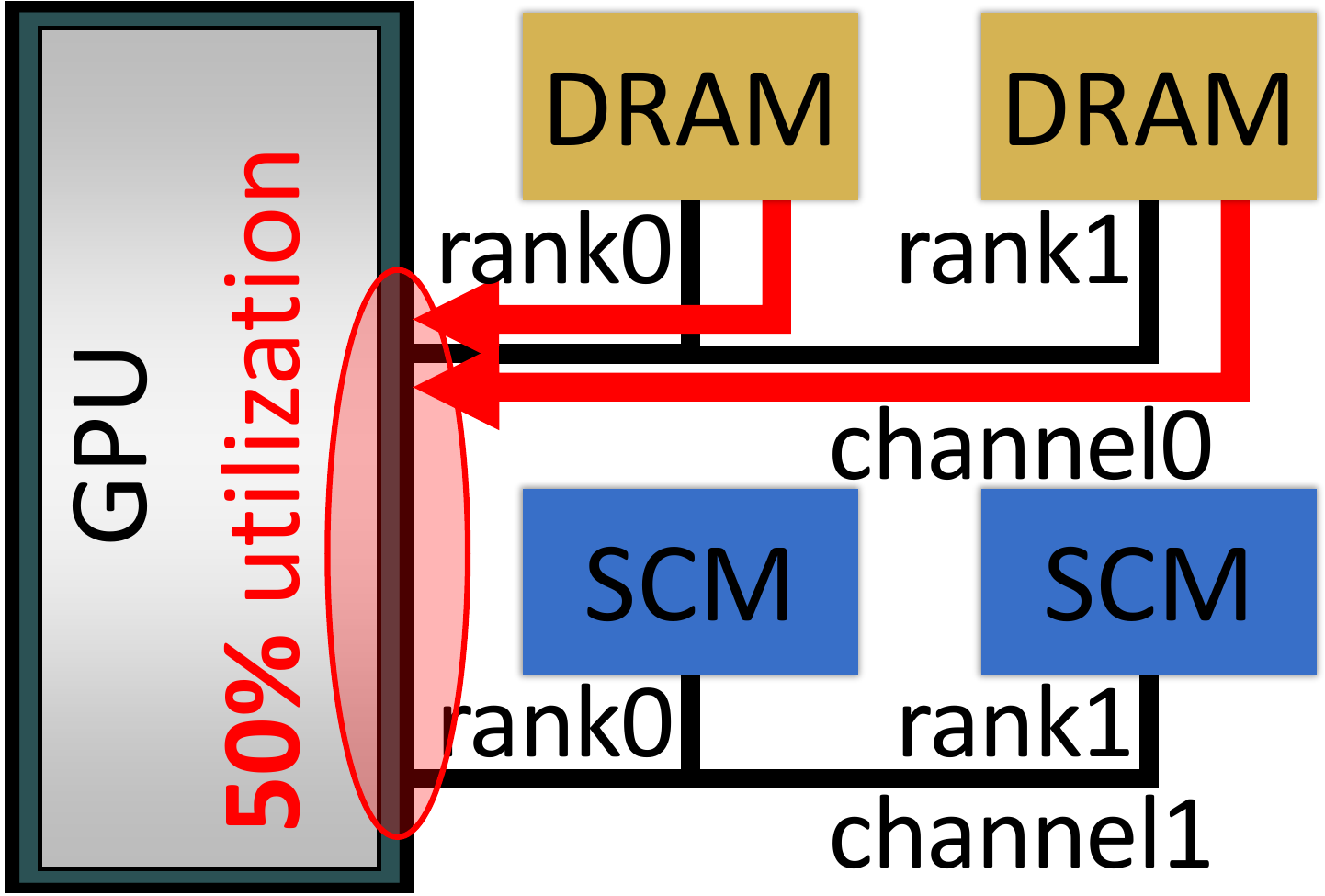}
\includegraphics[width=0.325\linewidth]{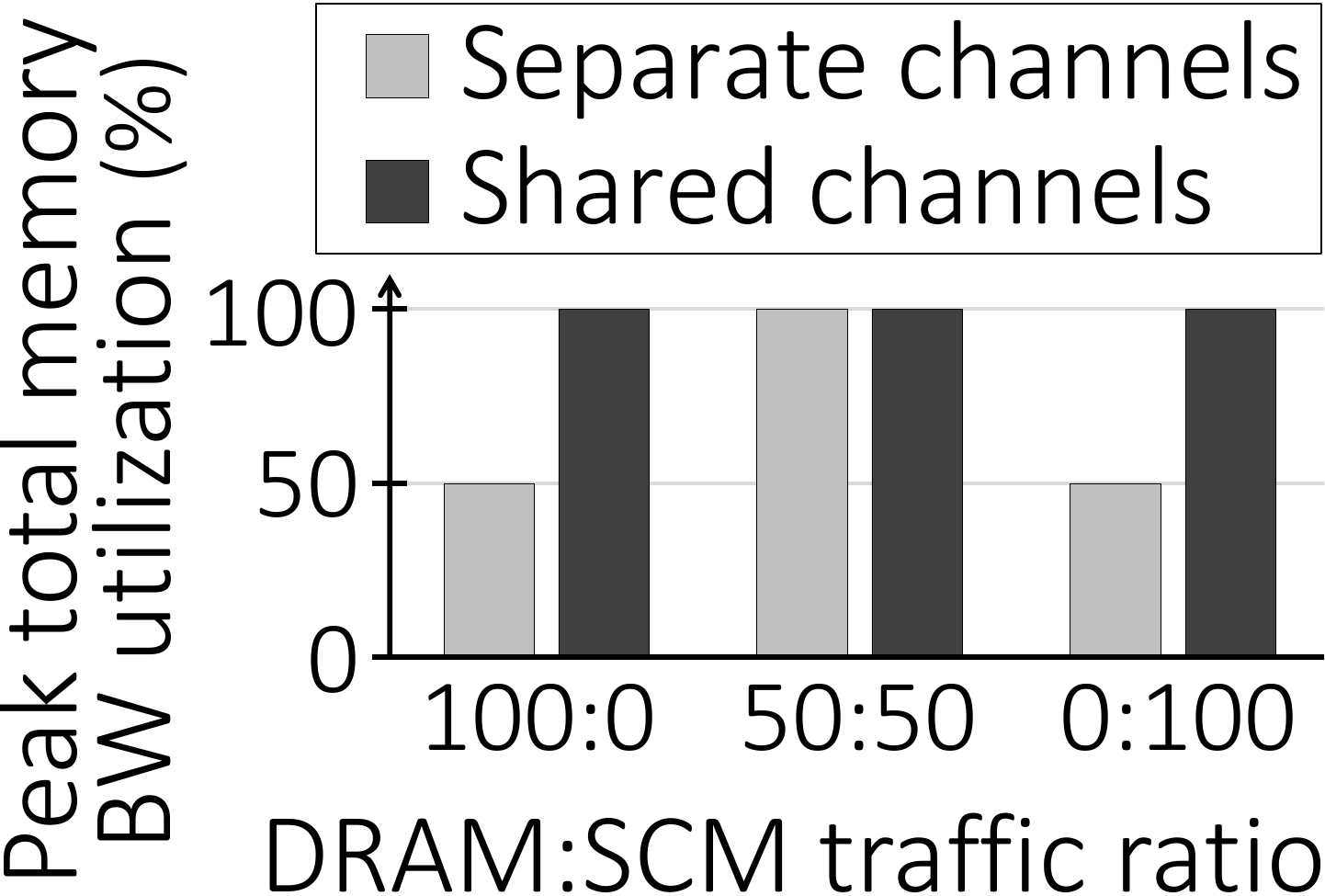} \\
\raggedright{\hspace{0.53in}\footnotesize{(a)} \hspace{.97in} \footnotesize{(b)}\hspace{1.0in} \footnotesize{(c)}}
\caption{Integration of SCM and DRAM cache using (a)  shared channels and (b) separate channels.
(c) Peak memory BW for varying DRAM-to-SCM traffic ratios.}
\label{fig:channel_vs_rank}
\end{figure}

To improve GPU's memory capacity under fixed pin BW, 
the DRAM cache and SCM can be integrated in a 3D-stack 
to create a Heterogeneous Memory Stack (HMS) shown in Fig.~\ref{fig:design_space}a
or as separate memory devices (Fig.~\ref{fig:design_space}b). 
The separate SCM devices can also be attached using external NVLink~\cite{nvlink}
or CXL~\cite{cxl}.
The designs differ in how the devices are mapped to memory channels.
In HMS, the DRAM and SCM can share the same channel as different ranks 
(Fig.~\ref{fig:channel_vs_rank}a) similar to~\cite{optane_eurosys22, tictoc},
whereas separate devices inevitably use separate buses 
(Fig.~\ref{fig:channel_vs_rank}b) similar to~\cite{knights_landing, accord, unisoncache, footprint_cache}.

However, for flexible channel BW utilization for varying traffic patterns, each channel should be shared by both the DRAM cache and SCM. For example, if a workload shows a high DRAM cache hit rate, the SCM-only channel in Fig.~\ref{fig:channel_vs_rank}b can become idle while the DRAM-only channel experiences high contention, resulting in only 50\% utilization overall. In contrast, in Fig.~\ref{fig:channel_vs_rank}a, both channels can be fully utilized by the DRAM caches in each channel (Fig.~\ref{fig:channel_vs_rank}c).
Optane DIMM is also placed on the same channel as DRAM DIMM~\cite{optane_micro20}.

Thus, we focus on the HMS design with DRAM and SCM ranks sharing the same channel 
implemented using TSVs in a 3D stack, but we show that our DRAM cache is also effective for
SCM integrated with separate channels (\S\ref{sec:perf}). 
HMS retains the HBM's high-level design and interface, including TSV connectivity, the number of banks and bank groups, and the base die's I/O buffers for signal integrity~\cite{hbm_arch}.
The key difference is replacing the upper-rank DRAM dies in HBM with SCM dies in HMS (Fig.~\ref{fig:design_space}a).
Thus, each channel has a DRAM cache rank and an SCM rank.
Although the DRAM cache is not addressable by the programmer, HMS has a larger addressable capacity than HBM due to SCM's higher bit density; HMS provides 2$\times$ addressable memory capacity than HBM, assuming SCM has 4$\times$ bit density compared to DRAM~\cite{pcm_isca09,3dxpoint}.
We use PCM as SCM due to their maturity~\cite{emerging_mem}, but other SCM devices can also be used.

Due to GPU's high memory BW demand, maximizing the effective BW is a key consideration for our DRAM cache.
HW-managed DRAM caches for CPUs typically use 64~B 
cachelines~\cite{loh_hill, alloycache, redcache, optane_basic}. 
In contrast, in this work,
we assume a 256~B DRAM cacheline\footnote{For L1 and L2 caches, we still assume 128~B line with 32~B sectors~\cite{accelsim}.} to achieve high memory bus utilization, amortize the long activation latency of SCM, and exploit the high spatial locality of memory accesses from GPUs. 
In addition, to reduce the BW overhead of fetching DRAM cache tags (hereafter, tags) and metadata (e.g., LRU bits), 
we make the DRAM cache direct-mapped and reduce the tag size. 
Combined with the large cacheline size, the small tag size enables capacity-effective on-chip tag caching (\S\ref{sec:tag_cache}) that further reduces the BW overhead of DRAM cache probes. 
To track DRAM cache misses, we use SRAM-based MSHR for each channel of the DRAM cache located near the memory controller.
DRAM cache operations (e.g., probe, fill, and eviction) are translated
by a DRAM cache controller into DRAM or SCM requests, which are then scheduled
by the memory controller, considering timing parameters.

\begin{figure}
\centering
\includegraphics[width=.98\columnwidth]{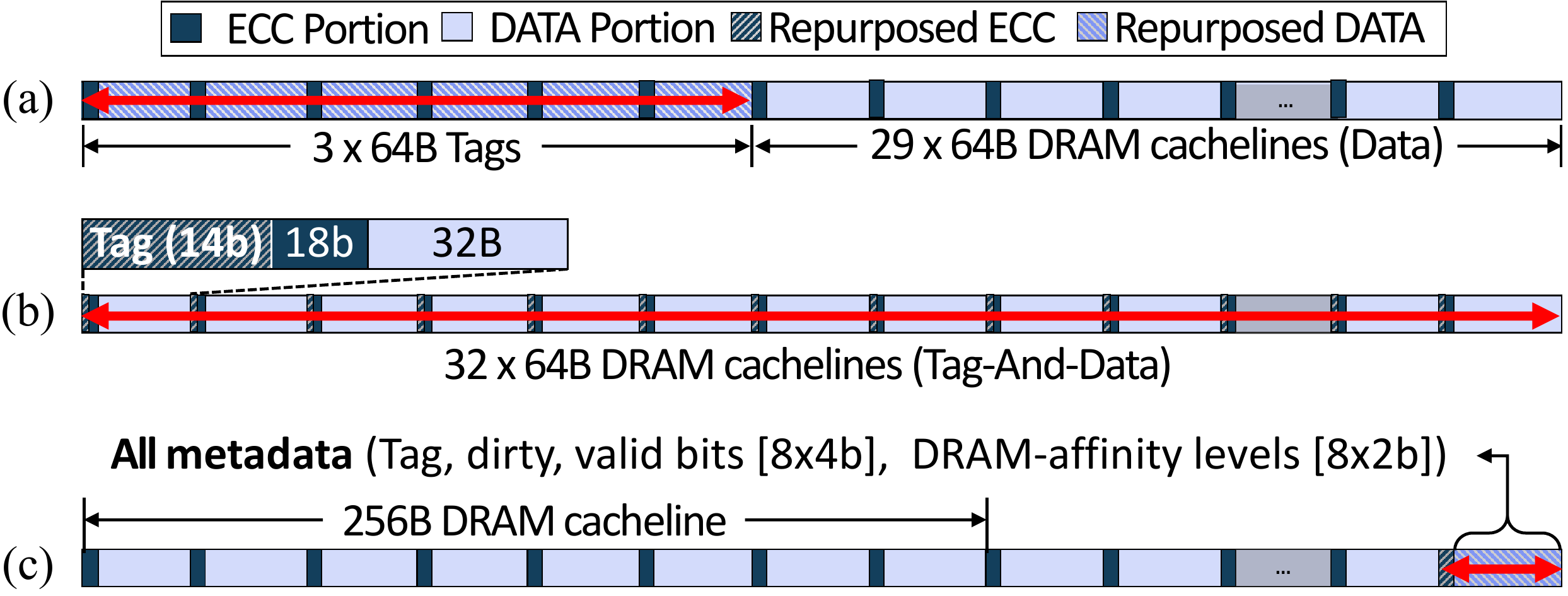}
\caption{DRAM cache row with (a) Loh-Hill cache~\cite{loh_hill}, 
(b) Tag-and-data (TAD)~\cite{alloycache, knights_landing}, and
(c) proposed Aggregated Metadata-In-Last-column (AMIL). 
The double arrows indicate the columns with tag information.} 
\label{fig:cacheline}
\end{figure}

\subsection{Aggregated Metadata-In-Last-Column (AMIL)}
\label{sec:amil}

Our CTC (\S\ref{sec:tag_cache}) keeps all tags of a DRAM cache row in a single L2 cache sector to exploit the high spatial locality of GPU workloads. 
Thus, we propose AMIL to minimize the tag access overhead by fetching all tags in a row with a single column access. 
Prior work on DRAM cache with conventional cacheline sizes~\cite{loh_hill} proposed a highly set-associative (e.g., 29-ways) organization that places tags in the first few columns and data in the remaining columns, requiring multiple columns to be accessed to fetch all tags in a row as shown with a double arrow in Fig.~\ref{fig:cacheline}a.
Alloy cache~\cite{alloycache} proposed a direct-mapped DRAM cache that fetches a Tag-And-Data (TAD) with a single access (Fig.~\ref{fig:cacheline}b).
However, TAD distributes the tags across all columns, so the entire row has to be accessed to fetch all tags. 
In addition, to comply with DRAM standards, DRAM caches 
with TAD~\cite{alloycache, bear, redcache, tictoc, accord, dice, carve}
have to repurpose some ECC bits to store tags,
degrading reliability.

To minimize BW overhead, AMIL places all metadata (tags, valid, dirty, and 
DRAM-affinity bits described in \S\ref{sec:affinity}) of a row 
in the last column's 32~B data portion (Fig.~\ref{fig:cacheline}c).
Although the last column cannot be used to cache SCM data, it accounts for a very small 
fraction (only 1.6\% for a 32~B column in a 2~KiB row) of a row.
Thus, AMIL effectively overcomes the reliability limitation of prior DRAM caches based on TAD.

The AMIL is enabled by the high DRAM/SCM capacity ratio and large cacheline size we propose,
unlike prior DRAM caches~\cite{loh_hill, alloycache, bear, tictoc, accord} with a few GiBs of DRAM cache for 10s-100s of GiBs of main memory.
Assuming SCM has 4$\times$ the capacity of a DRAM die and using a direct-mapped DRAM cache, the DRAM cache tag is 2-bit. With valid/dirty bits and 2-bit DRAM-affinity, each cacheline only requires 6-bit metadata. 
With the 256~B DRAM cacheline and a 2~KiB row, each row includes 8 cachelines, needing only 48 bits for metadata. This metadata is also
protected with ECC.

\subsection{SCM-aware DRAM Cache Bypass Policy}
\label{sec:bypass}
Our SCM-aware DRAM cache bypass policy considers the multi-dimensional characteristics of accesses (\S\ref{sec:dram_cache_motivation}),
i.e., spatial locality, hotness, and write intensity,
to keep useful data in DRAM and avoid DRAM cache thrashing from 100,000s of GPU threads.
The key insight is that we can quantify the combined effects of these three-dimensional characteristics with a one-dimensional score metric. 
First, our \emph{SCM penalty score} accounts for the spatial locality and write intensity by comparing the latency penalty of SCM versus DRAM for given requests.
Then, the score is multiplied by hotness (i.e., per-page activation counter) to obtain the final \emph{DRAM-affinity score}. The scores can be calculated during runtime
at a low cost (\S\ref{sec:putting_it_all_together}), 
without any separate profiling phase.

\subsubsection{SCM Penalty Score}
\label{sec:dev_sensitivity}
The SCM penalty score reflects the \emph{latency penalty of SCM per column access}. High-penalty accesses cache data in DRAM, while others bypass the DRAM cache to access SCM directly.
The score considers the spatial locality within the row buffer and differentiates writes from reads.
The spatial locality is essential to consider, as memory accesses with many row buffer hits can amortize the long SCM activation latency. Consequently, accessing such data from SCM has a lower performance impact than when few row buffer hits occur.
On the other hand, write-intensive data should be cached in DRAM because 
the write latency is higher for SCM than for DRAM and SCMs can have limited write endurance~\cite{scm_guideline}.

For the bypassing decision, the latency of memory accesses to an SCM row is first calculated based on timing parameters for required operations such as row activation, column accesses, write recovery, and precharge. Similarly, the latency required to serve the same memory accesses from DRAM (i.e., as if they were all accessed from DRAM) is also calculated.
The difference between these two latencies is then divided by the number of column accesses to obtain SCM's per-access penalty. 

\begingroup\makeatletter\def\f@size{9.5}\check@mathfonts
\begin{equation} \label{eq:1}
    {SCMPenaltyScore}=\frac{Latency_{SCM} - Latency_{DRAM}}{NumColumnsAccessed}
\end{equation}
\endgroup

\begin{figure}
\centering
\includegraphics[width=1.0\linewidth]{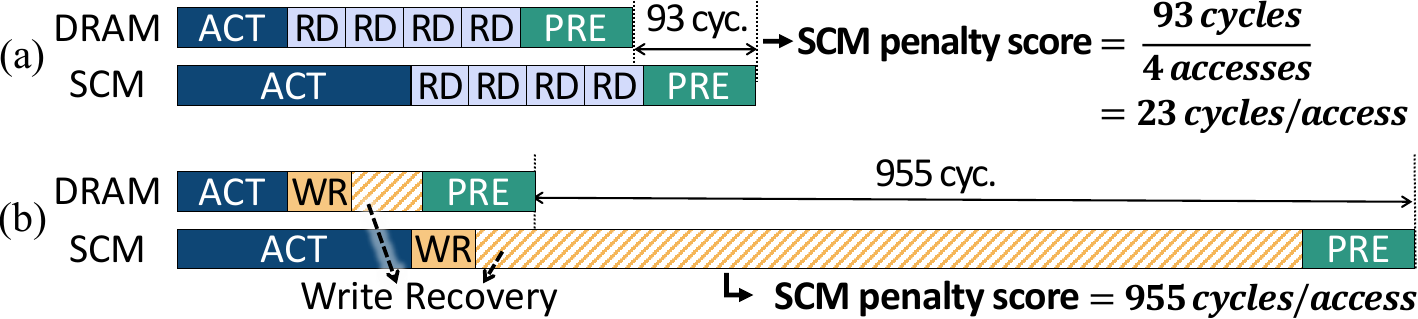}\\
\caption{Example timing diagrams contrasting the SCM penalty score calculated
when a row is accessed with (a) multiple read accesses and (b) a single write access.
Not drawn to scale.}
\label{fig:timing_example}
\end{figure}

For example, when there are no writes and SCM's long activation delay  
is well amortized over multiple column accesses, the SCM penalty score is low (Fig.~\ref{fig:timing_example}a).
In contrast, when there is a write without spatial locality, the latency discrepancy between SCM and DRAM is large, and the SCM penalty score is high (Fig.~\ref{fig:timing_example}b). 
With the scores, our policy (\S\ref{sec:bypass_policy_sub}) can bypass the DRAM cache for the access pattern in Fig.~\ref{fig:timing_example}a and cache data accessed in Fig.~\ref{fig:timing_example}b. Thus, DRAM contention can be reduced while keeping data in DRAM when it is beneficial.

The SCM penalty score can be computed at a low cost. Because column access latency is
identical between SCM and DRAM~\cite{scm_guideline},
it will be canceled out in the numerator of Eq.~\ref{eq:1}.
Thus, the numerator can be approximated and statically pre-computed as $(t_{RCD,SCM} - t_{RCD,DRAM})$ 
if the accesses include only reads or as $(t_{RCD,SCM} - t_{RCD,DRAM} + t_{WR,SCM} - t_{WR,DRAM})$
if writes are included. Then, it is simply divided by the number of columns accessed, which can be
recorded in the DRAM cache's MSHR along with the presence of write. This implementation requires
two 32-bit registers for the pre-computed values and an ALU.

\subsubsection{DRAM-Affinity Score}
\label{sec:affinity}
The SCM penalty score incorporates the spatial locality and presence of writes 
but ignores data's access frequency, which requires historical information 
and cannot be inferred from current requests.
Thus, we propose another score metric called \emph{DRAM-affinity score}, calculated by multiplying a request's SCM penalty score with its per-page \emph{activation counter}.
The activation counter is incremented when a DRAM or SCM row is activated.
The DRAM-affinity score is discretized into $N_{levels}$ levels
with a fixed interval
and kept in the DRAM cache as metadata (Fig.~\ref{fig:cacheline}(c)) for bypass policy.

\subsubsection{SCM-aware DRAM Cache Bypass Policy}
\label{sec:bypass_policy_sub}

\begin{figure}
\centering
\includegraphics[width=0.90\linewidth]{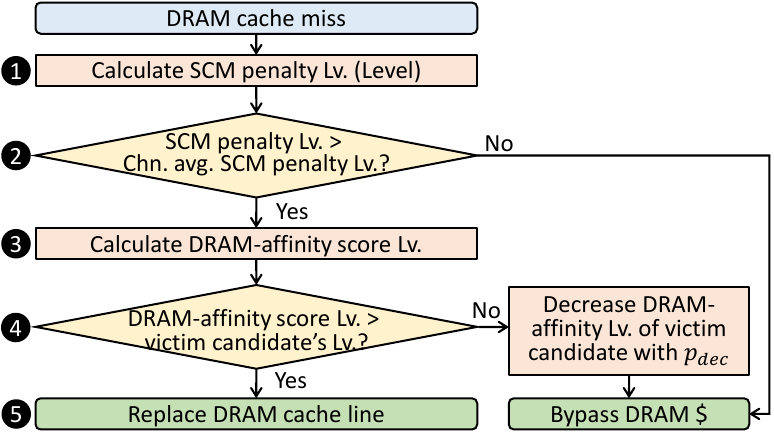}
\caption{SCM-aware DRAM cache bypass policy.}
\label{fig:flow_chart}
\end{figure}

Because accessing the victim DRAM cacheline's DRAM-affinity level for every DRAM cache miss would 
incur very high BW overhead, we propose a two-level bypass policy to minimize this overhead.
The first-level comparison is done to filter the majority of the requests without any DRAM BW overhead. 
If the comparison is passed, the second-level comparison is done using the victim's metadata in DRAM. 

First, as shown in Fig.~\ref{fig:flow_chart}, when a DRAM cache miss occurs, 
the SCM penalty score for the requests mapped to the same row is calculated and discretized to $N_{levels}$ levels between 0 and the maximum value observed so far (\circled{1}).
This discretization prevents inconsistent bypass decisions due to small fluctuations in the score.
The discretized score is then compared to a similarly-discretized moving average of the SCM penalty score maintained by the memory controller (\circled{2}). 
If the request's score level is less than or equal to the average level, the DRAM cache is bypassed
(i.e., no miss fill is done).

Otherwise, the current request's DRAM-affinity level is compared with the victim cacheline's DRAM-affinity level
(\circled{3}, \circled{4}). 
If the current request has a higher level, the victim is replaced (\circled{5}), and the affinity level is stored. 
If the victim is invalid, the miss fill is done without this comparison. 
If the replacement is not done, the victim's score level is decremented with a probability $p_{dec}$ to adapt to changing working set.
$p_{dec}$ is calculated as the accessed page's activation counter divided by the maximum activation counter observed by this memory controller.
The intuition is that the victim's DRAM-affinity level should be more likely to be decremented if hot data bypassed the DRAM cache.
Score calculations can be done with an FPU with six 32-bit registers to hold
average, maximum, and current request values for SCM-penalty score and DRAM-affinity scores.
The activation counters can be tracked in 2~MiB granularity with low overhead
as 160~GiB GPU memory requires only 80~KiB from 80-kilo entries of 8 bits counters.
To address counter saturation, a 3-bit register indicates the position of the MSB bit to implement
low-cost right-shifts. The LSB bits can be zeroed over time and ignored until the shifts are finished. 
Depending on the workload's characteristics, the activation counters can be used selectively
(e.g., use a constant value of 1 if its benefit is not high).

\begin{figure}
\centering
\includegraphics[width=0.90\linewidth]{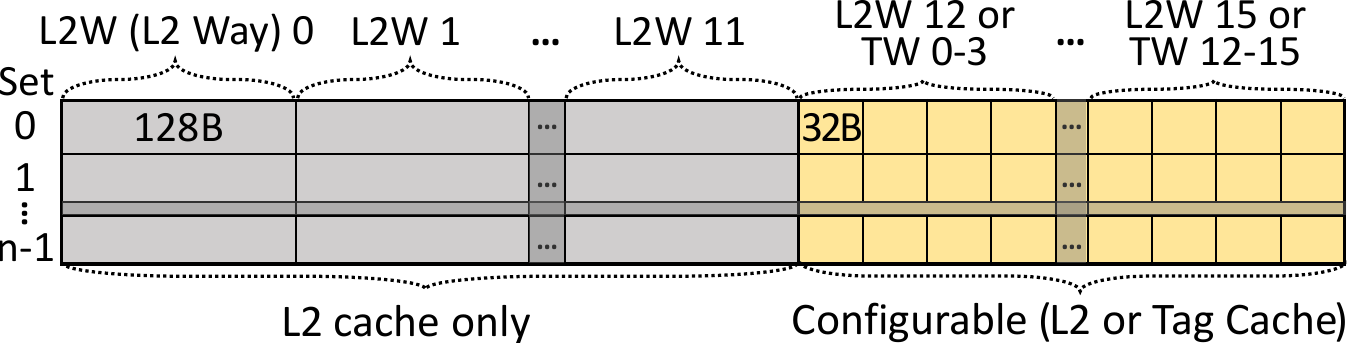}
\caption{Configurable Tag Cache in L2 cache, assuming the Tag Cache (TC) can use up to 4 L2 cache ways and each L2 cache way can hold 4 TC ways.}
\label{fig:tag_cache}
\end{figure}

\subsection{Configurable Tag Cache (CTC)}
\label{sec:tag_cache}

Our bypass policy reduces DRAM traffic for DRAM cache misses. 
However, determining a hit or miss also requires a tag access from DRAM~\cite{alloycache} for every
L2 cache miss, incurring high overhead. 
Using an additional on-chip \emph{tag cache} to hold DRAM cacheline tags~\cite{bear}
can incur high overhead for our DRAM cache, due to its significantly larger capacity. 
The MissMap~\cite{loh_hill} approach statically partitions the LLC to hold 
only DRAM cache presence information with low overhead, instead of the full tags.
However, such static partitioning can degrade the performance when high LLC capacity is required. For example, recent GPUs support L2 cache resident control, whereby programmers can specify some data to persist in the L2 cache for high performance~\cite{a100_gtc}.
To meet the varying demands of workloads, flexible partitioning is necessary.

Thus, we propose a \emph{Configurable Tag Cache (CTC)} to reduce the DRAM cache probe traffic and support flexible partitioning between the L2 cache and tag cache without a separate SRAM for caching tags (Fig.~\ref{fig:tag_cache}).
The programmer can specify the number of tag cache ways out of the total L2 cache ways, similar to how the user chooses the split between the L1 data cache and shared memory~\cite{a100}.
In general, workloads with high DRAM BW demand can benefit more from additional CTC ways, as a CTC miss generates DRAM cache probe traffic that contends with the demand DRAM traffic. 
The number of CTC ways can also be determined by profiling~\cite{shm_profiling}
or set-dueling~\cite{set-dueling}.
For iterative workloads~\cite{parboil,rodinia,graphbig}, it can be changed across kernels by flushing dirty lines, but we leave such a study for future work. 
If DRAM is configured as part of memory~\cite{knights_landing}, all ways are used for L2 cache.

A single L2 cache way is divided into four 32~B Tag Cache ways, assuming a 128~B L2 cacheline.
The size of the DRAM tags for a row is 4~B, excluding the DRAM-affinity bits not kept in CTC (\S\ref{sec:amil}). Thus, a Tag Cache line is further divided into eight 4~B sectors that are mapped to eight DRAM rows. 
To minimize area overhead, we assume that up to four L2 cache ways can be used for tag caching. 
The CTC requires modification of L2, but its overhead is low as it adds only 8+8+22=38 bits (per-sector valid and dirty bits, and per-line tag) per cacheline and 4-bit pseudo-LRU metadata per set.
The storage overhead is 612 bits per set or only 2.5\% of L2 cache.

\subsection{Power Management and Performance Optimization}
\label{sec:opt}

Accessing SCM cells can require higher energy and power consumption than DRAM,
leading to higher temperatures. Especially, SCM power consumption needs to be
managed for HMS that stacks SCM on DRAM.  Thus, we propose a simple 
SCM power throttling technique that monitors the memory stack's temperature~\cite{hbm} 
and adjusts SCM's timing parameters.
If the temperature increases too high, the timing parameters for SCM activation
($t_{RCD}$) and/or write recovery ($t_{WR}$) are doubled to limit power consumption.
In our evaluation, throttling is rarely required, but it can effectively
curtail SCM's power and temperature increase if needed.

In addition, when the memory footprint is small (e.g., based on the
memory allocation for UM), GPU's DRAM can be used as part of memory along with SCM,
rather than a cache. For high performance, data can initially be placed in DRAM, with the remaining data mapped to SCM. Additionally, SCM can operate in SLC mode, instead of MLC mode, for enhanced performance.
As a result, the GPU can minimize performance impact for small workloads
and our evaluation results show that HMS can provide high performance for
varying memory footprints. 

\subsection{Putting It All Together}
\label{sec:putting_it_all_together}

The operations of our DRAM cache can be summarized as follows.
When an L2 cache miss occurs, the CTC is first looked up.
If a CTC hit, it is immediately determined whether the request hits the DRAM cache. If not, the DRAM cache must be probed to access the tag and fill CTC.  
With AMIL, tags for the entire row are fetched with a single DRAM access, amortizing the probe overhead for subsequent accesses.
If a DRAM cache hit occurs, the request accesses DRAM and the average SCM penalty score is updated.
Otherwise, the requested address is first accessed from SCM to serve the demand access,
and then, the DRAM cache bypass policy (\S\ref{sec:bypass_policy_sub}) determines if the DRAM cache fill should be done.
With 128~B L2 cacheline and 32~B sectors, all L2 fills are done in 32~B
size whether it is fetched from DRAM cache or SCM, 
whereas data movement between DRAM and SCM uses 256~B DRAM cacheline size.

\section{Evaluation}
\label{sec:eval}

\subsection{Methodology}
\label{sec:methodology}

\begin{table}
\caption{Simulated system configuration.}
\scriptsize
\setlength{\tabcolsep}{1.8pt} %
\begin{center}
    \renewcommand{\arraystretch}{1.00}
    \begin{tabular}{|l|c|}
    \hline
    \multicolumn{2}{|c|}{\textbf{SMs}} \\ 
    \hline
    \multicolumn{2}{|c|}{21 SMs, 64 warps/SM, 65536 regs/SM, clock frequency: 901~MHz} \\ 
    \multicolumn{2}{|c|}{L1+shared memory: 192~KiB/SM, 128~B line (32~B sectors), LRU} \\
    \multicolumn{2}{|c|}{L1 SRAM latency and BW: 15~cycles and 17~GB/s/SM} \\
    \hline
    \hline
    \multicolumn{2}{|c|}{\textbf{L2 cache and CTC parameters}} \\
    \hline
    \multicolumn{2}{|c|}{L2(Baseline): 128~B line (32~B sectors), 16 ways, 8~MiB capacity, LRU} \\
    \multicolumn{2}{|c|}{L2(HMS): 128~B line (32~B sectors), 12 ways, 6~MiB capacity, LRU} \\
    \multicolumn{2}{|c|}{CTC(HMS): 32~B line (4~B sectors), 16 ways, up to 2~MiB capacity, LRU} \\
    \multicolumn{2}{|c|}{Freq: 901MHz, latency:120~cycles, peak BW: 402GB/s from 16 banks} \\
    \hline
    \hline
    \multicolumn{2}{|c|}{\textbf{Memory organization (for both DRAM and SCM)}} \\
    \hline
    \multicolumn{2}{|c|}{row buffer: 2 KiB, bus width: 128 bit (BL 2, DDR), \# of channels: 8, \# of dies: 8,} \\ 
    \multicolumn{2}{|c|}{\# of bank groups per ch.: 4, \# of banks per bank group: 4, FR-FCFS scheduler} \\
    \multicolumn{2}{|c|}{Bus frequency: 1~GHz, Bus peak BW: 256~GB/s from 8 channels} \\
    \hline
    \hline
    \multicolumn{2}{|c|}{\textbf{Timing parameters}} \\
    \hline
    DRAM~\cite{ramulator} & CL: 14, RCD: 14, RAS: 33, WR: 16, RP: 14 \\
     & (row hit:15ns, row miss(closed page):43ns) \\
    \hline
    SCM~\cite{scm_guideline, pcm_hynix}  & CL: 14, RCD: 120, RAS: 120, WR: 1000, RP: 14 \\
     & (row hit:15ns, row miss(closed page):149ns) \\
    \hline
    \hline
    \multicolumn{2}{|c|}{\textbf{Unified Memory-related latency and BW}} \\
    \hline
    \hline
    PCIe link & BW: \textbf{12.8~GB/s} (i.e., 1/5 of PCIe 4.0 $\times$16) or 64~GB/s (\S \ref{sec:sensitivity_study}) \\
    \hline
    NVLink & Latency for CPU memory access (cacheline size): 0.135~$\mu$s~\cite{adaptive} \\
    (where applicable) & BW: 76.8~GB/s (CPU memory BW of 46.6~GB/s) \\
    \hline
    Other & Page fault handling latency: 20$\mu$s~\cite{batchaware} \\
    \hline
    \hline
    \multicolumn{2}{|c|}{\textbf{Memory energy (pJ/bit)~\cite{yavits2020wolfram, lee2009architecting}}} \\
    \hline
    DRAM & ACT: 1.17, PRE: 0.39, RD: 0.93, WR: 1.02 \\
    \hline
    SCM & ACT: 2.47, PRE (WR): 16.82, RD: 0.93, WR: 1.02 \\
    \hline
    \end{tabular}
\end{center}
\label{tab:sim_config}
\end{table}

\begin{figure*}
\centering
\includegraphics[width=\linewidth]{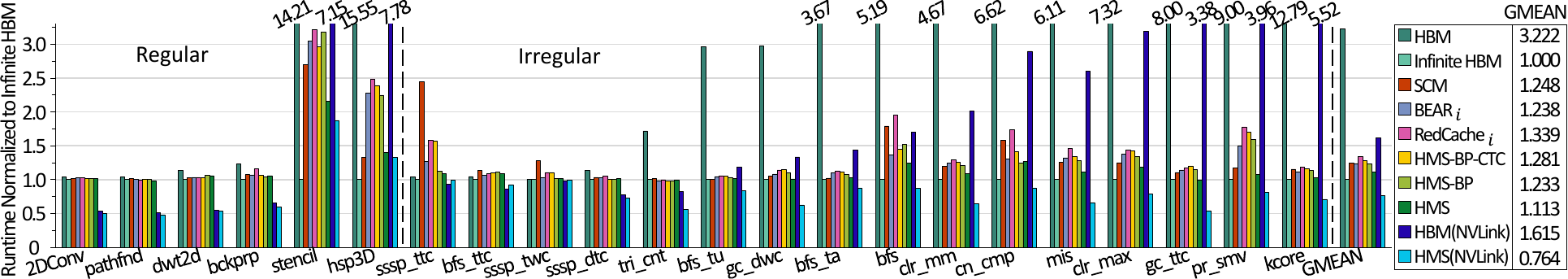}
\caption{Runtime of GPUs with different memory designs normalized to Infinite HBM. PCIe was used for host connectivity unless otherwise stated.}
\label{fig:perf_75}
\end{figure*}

We integrated Accel-sim~\cite{accelsim} with a UM model~\cite{uvmsmart} and Ramulator~\cite{ramulator} for simulation (Table~\ref{tab:sim_config}). 
Due to the very long simulation time of the oversubscribed baseline 
(\S\ref{sec:um_model}), we downscaled an NVIDIA A100 GPU by 1/5 while keeping constant ratios between SM count, L2 cache capacity, memory channel count, and PCIe 
 (or NVLink) lanes.\footnote{Simulating a single workload took up to 24 days even with the downscaling. To simulate a full A100 GPU, the workloads' problem sizes needed to be scaled up accordingly to prevent a significant portion of the memory footprint from fitting in the 40~MiB L2 cache, which would substantially increase the simulation time. }
In addition, we also show results using the full 64~GB/s PCIe BW (\S\ref{sec:sensitivity_study}).
We used AccelWattch~\cite{accelwattch} to model GPU energy and 8~pJ/bit PCIe or NVLink energy~\cite{off-chip-energy}.

We used 22 workloads~\cite{graphbig, pannotia, parboil, rodinia, polybench} with memory footprints ranging from 19 to 135 MiB (68 MiB on average), excluding those with smaller footprints. 
We define R$_{HBM}$ as the relative capacity of HBM compared to the memory footprint and 
assume R$_{HBM}$=75\% (i.e., HBM holds 75\% of the workload's memory footprint) unless otherwise stated.
To model oversubscription, we adjusted HBM's capacity (i.e., the number of page frames available) as in all prior works~\cite{batchaware, uvmframework, uvmsmart, gpudmm, 2016:hpca:zheng} for simulation feasibility (\S\ref{sec:um_model}).
Other memory stacks were also configured to have the same capacity per DRAM die, and 4$\times$ capacity per SCM die compared to a DRAM die.
For instance, for a 100~MiB workload, HBM has a 75~MiB capacity while the DRAM cache and SCM have 37.5~MiB and 150~MiB capacities, respectively.
We also evaluated SCM-only 3D-stack (``SCM'') and an ideal HBM (``Infinite HBM'' or ``InfHBM'') with unlimited capacity (i.e., never oversubscribed). 
The SCM timing parameters we assume are conservative, considering real SCM device~\cite{hynix_pcm_256gb} has demonstrated shorter latencies.
The SCM energy parameters are also conservative as we assume a higher energy 
than the energy reported in a recent study of SCM~\cite{stern2021uncovering}.

For our DRAM cache, we focus on HMS due to its high speedups but also present results
with separate DRAM/SCM buses (\S\ref{sec:perf}).
We assumed $F_{update}$=100 and $N_{levels}$=4 for HMS. 
For the moving average, a new value has a weight of 1\%.
We disabled the activation counter for simplicity although an ideal activation counter's speedup is up to 7.6\% (0.4\% overall). 
To understand the impact of each technique, we also evaluated HMS without bypass and CTC (HMS-BP-CTC or HMS-B-C) as well as HMS without bypass (HMS-BP or HMS-B).
For conservative evaluation of CTC, its size was reduced to hold only a quarter of the total tags in the DRAM cache and ranged between 1-4~KiB across workloads.
The total DRAM cache tags of a full A100 GPU that replaces 40~GiB HBMs with equivalent HMSes is 40~MiB -- equal to the L2 cache capacity. 
Thus, we configure the CTC to use a quarter of the 16 L2 ways to 
hold a quarter of all DRAM cache tags.

We also evaluated prior works on BW-efficient DRAM caches (with 64~B DRAM cachelines) 
adopted for the DRAM cache within HMS.
For BEAR~\cite{bear}, we modeled an ideal DRAM Cache Presence bit such that the DRAM cache presence is known without LLC lookup or DRAM cache probe overhead, and refer to it as BEAR$_{i}$; for its Neighboring Tag Cache, we assumed the same 704~B/channel as in~\cite{bear}.
For RedCache~\cite{redcache}, we assumed an ideal gamma update without DRAM BW overhead and refer to it as RedCache$_i$.
For the mostly-clean DRAM cache~\cite{mostly_clean}, we assumed a perfect cache predictor and zero-cost tag probes, referring to it as McCache$_i$.

We assumed input data were initially in host memory, and we used the TBN prefetcher and pre-eviction policies for UM~\cite{uvmsmart} (\S \ref{sec:um}), which migrate data in 4~KiB to 1~MiB granularity adaptively, as in NVIDIA GPUs.\footnote{Using a first-touch policy instead of the NVIDIA UM scheme significantly degraded performance by 2.75$\times$ overall.}
We also studied replacing PCIe with high-BW NVLink for host connectivity. We kept the BW ratios of CPU/GPU memory and NVLink the same as in NVIDIA Grace Hopper Superchip~\cite{gh100} (Table \ref{tab:sim_config}). We modeled the dynamic access counter scheme for NVLink, which considers
the amount of free memory capacity and access frequency to migrate hot pages 
to the GPU while cold data is accessed directly from the remote memory in cacheline granularity~\cite{um, adaptive}.
For several plots, we only show representative workloads due to space constraints, but
the average values reported are always calculated over all workloads.

\subsection{Performance}
\label{sec:perf}
\begin{figure}
\centering
\includegraphics[width=0.95\linewidth]{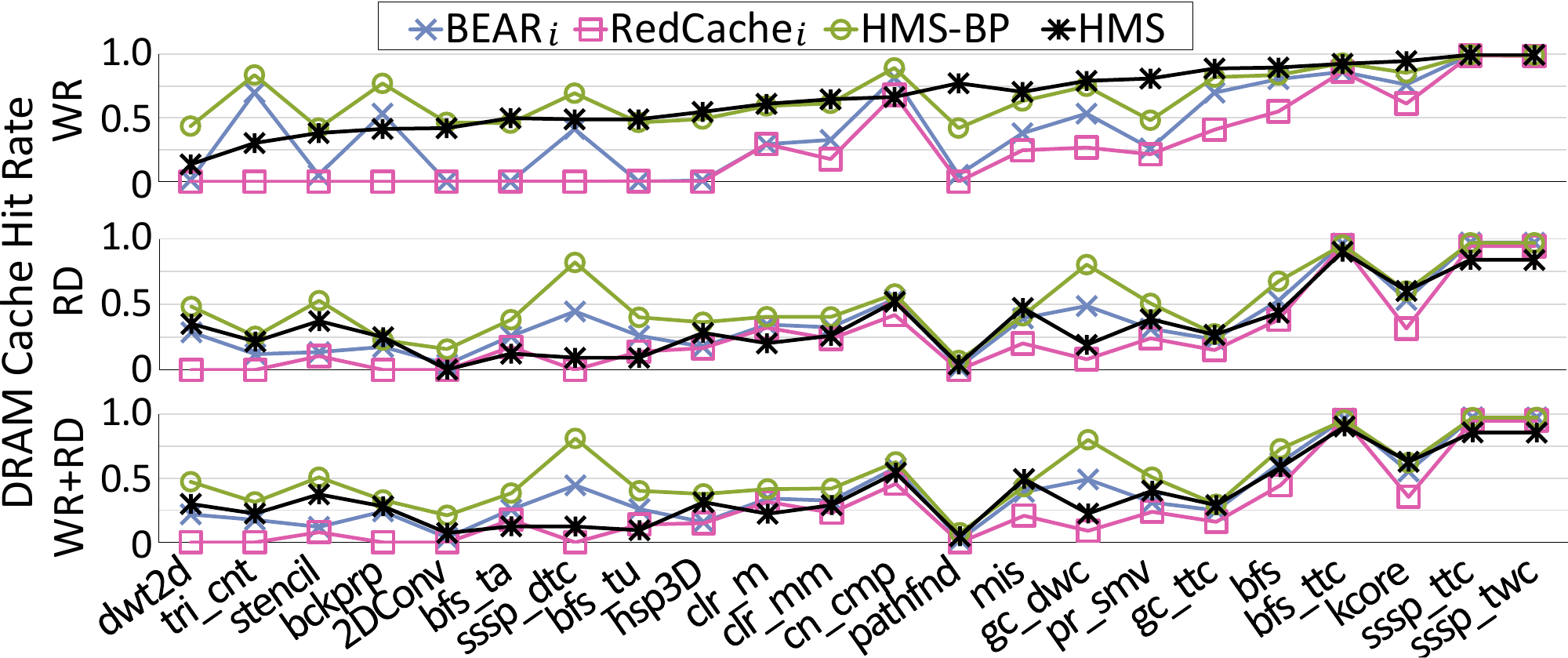}
\caption{DRAM cache hit rates with different designs.}
\vspace{-.1in}
\label{fig:hit_rate}
\end{figure}

Compared to the oversubscribed HBM, HMS can hold the entire memory footprint and achieved a significant speedup of up to 12.5$\times$ (2.9$\times$ overall) by reducing data transfers over PCIe by up to 159$\times$ for \textfw{stencil} (7.3$\times$ on average) (Fig.\ref{fig:perf_75}).
The speedup was especially pronounced for graph workloads with irregular access patterns, for which UM page prefetchers are ineffective.
Despite having a smaller DRAM capacity than HBM, our DRAM cache effectively filters out requests to SCM.
For example, SCM resulted in up to 2.25$\times$ longer runtime than HBM for \textfw{sssp\_ttc}, as the SCM was frequently accessed with little row buffer locality for writes. In contrast, HMS reduced its performance impact using the
DRAM cache with write hit rates of 99.6\% (Fig.~\ref{fig:hit_rate}).
Because \textfw{sssp\_ttc} has a relatively smaller working set per kernel, 
it did not suffer significantly from oversubscription with HBM. 
For some graph workloads (e.g., \textfw{bfs\_tu}, \textfw{bfs\_ta}, \textfw{gc\_*}, \textfw{clr\_*}, etc.) with a relatively higher row buffer locality and/or low write-intensity, DRAM cache hit rates for HMS were relatively low at 10-30\% due to bypass, but write requests still had high hit rates of 49-89\%.
For some workloads with high row buffer locality and read-intensity, SCM achieved similar performance as InfHBM as the long activation latency was amortized.

For regular workloads, HBM's performance varied depending on the working set size and the effectiveness of the UM prefetcher. While it had similar performance as InfHBM for some workloads (e.g., \textfw{pathfnd} and \textfw{2DConv}), it suffered significantly for others (e.g., \textfw{stencil} and \textfw{hsp3D}).
In contrast, HMS reduced the performance gap between the HBM and InfHBM from 15.55$\times$ (14.21$\times$) to 1.40$\times$ (2.15$\times$) for \textfw{hsp3D} (\textfw{stencil}) with higher capacity.
Overall, HMS outperformed HBM and SCM by 2.9$\times$ and 12.1\% on average, respectively, achieving within 11.3\% of the performance of the InfHBM.

\begin{figure}
\centering
\includegraphics[width=0.96\linewidth]{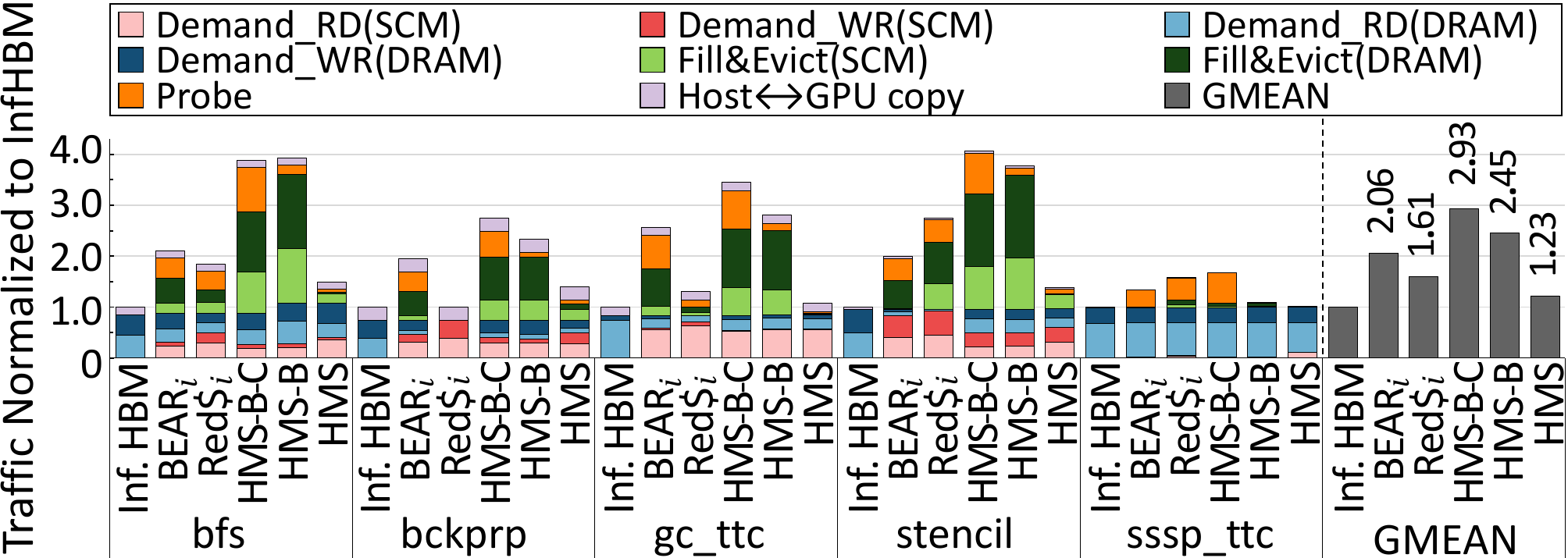}
\caption{Traffic breakdown of DRAM cache designs relative to InfHBM.} %
\label{fig:traffic_breakdown}
\end{figure}

\begin{figure}
\centering
\includegraphics[width=0.96\linewidth]{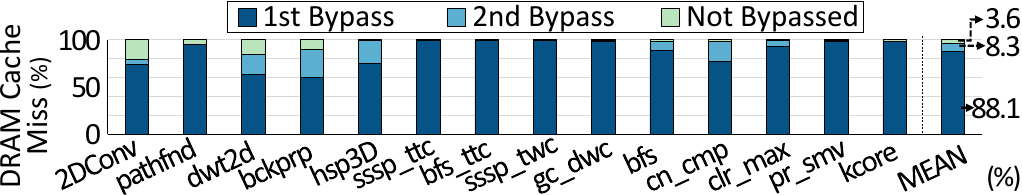}
\caption{DRAM cache bypass breakdown.} %
\label{fig:bypass_breakdown}
\end{figure}

\noindent
{\bf \textit{Impact of bypass and CTC.}}
Disabling bypass (HMS-BP vs. HMS) increased DRAM writes by 5.5$\times$ and SCM writes by 3.2$\times$ for write-backs, resulting in 2.4$\times$ more memory traffic overhead than InfHBM because all DRAM cache misses caused 256~B cacheline fills.
As a result, runtime was increased by up to 60\% for \textfw{hsp3D} (10.8\% overall). 
However, enabling the bypass reduced the traffic overhead to only 1.23$\times$ (Fig.~\ref{fig:traffic_breakdown}), reducing DRAM and SCM demand access latencies by 58.5\% and 27.2\%, respectively. 
Most bypasses (88.1\%) were done with the first comparison using the SCM penalty level without accessing the DRAM-affinity level of the victim in DRAM (Fig.~\ref{fig:bypass_breakdown}).
Nevertheless, the second comparison is essential in preventing evictions by cachelines with a smaller or equal DRAM-affinity level.
Disabling it increased runtime by up to 49\% for \textfw{stencil} (4.8\% overall).
Compared to HMS-BP-CTC, enabling CTC (i.e., HMS-BP) provided speedups of up to 40\% (3.9\% overall), thanks to high CTC hit rates of 91\% overall (59\% at minimum), which reduced DRAM probes.
CTC reduced memory traffic overhead over the InfHBM from 2.93$\times$ to 2.45$\times$ (Fig.~\ref{fig:traffic_breakdown}), and DRAM demand access latency by 45\%. 
Reserving four L2 ways for CTC only had 0.9\% impact overall over an ideal full L2 cache with zero-cost CTC.

\noindent
{\bf \textit{Comparison to prior work.}}
HMS outperformed BEAR$_{i}$ and RedCache$_{i}$ by up to 62.0\% (11.2\% overall) and 77.1\% (20.2\% overall), respectively, as these designs did not consider SCM's low performance.  
Thus, they had very low DRAM cache write hit rates overall than HMS (Fig.~\ref{fig:hit_rate}),
resulting in higher demand SCM write traffic -- e.g.,  
1.76$\times$ (3.97$\times$) for \textfw{bfs} with Bear$_{i}$ (RedCache$_{i}$).
In particular, RedCache$_{i}$ had zero DRAM cache hits for several workloads as it bypassed DRAM caching for pages with low access counts.
With CTC, HMS also reduced DRAM cache probe traffic by 93.1\% (90.6\%) overall compared to BEAR$_{i}$ (RedCache$_{i}$). 
Although CTC increased L2 miss rate by 5.4\%, overall memory traffic of HMS was 40.5\% 
(23.6\%) lower than that of BEAR$_{i}$ (RedCache$_{i}$).
McCache$_{i}$ showed high SCM write traffic 
(1.85$\times$ more than HMS overall) due to its partial write-through DRAM cache and lack of SCM-awareness, and underperformed BEAR$_{i}$.

\begin{figure}
\minipage{0.49\linewidth}
\includegraphics[width=\linewidth]{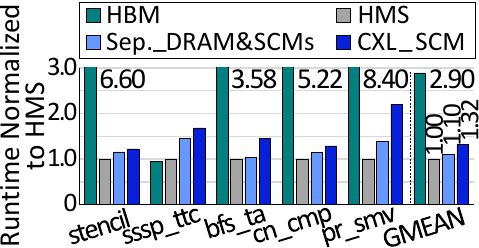}
\endminipage\hfill
\minipage{0.49\linewidth}
\includegraphics[width=\linewidth]{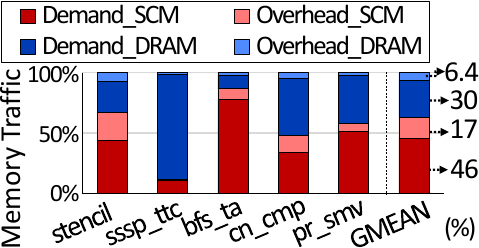}
\endminipage
\vspace{.09in}
\raggedleft{\hspace{1.7in}\footnotesize{(a)} \hspace{1.5in} \footnotesize{(b)}\hspace{0.7in}}
\caption{(a) Performance of alternative integration of our DRAM cache and SCM. (b) Memory traffic breakdown of HMS.}
\label{fig:memory_traffic_breakdown}
\end{figure}

\noindent
{\bf \textit{HMS design space exploration.}}
We also evaluated alternative integration of DRAM and SCM, including CXL interface.
For CXL, we assumed GPU used integrated CPU cores~\cite{mi300apu} as CXL host.
Using separate DRAM and SCM devices (``Sep.\_DRAM\&SCM'') or CXL-attached SCM (``CXL\_SCM'') still outperformed HBM by 2.6$\times$ and 2.2$\times$, respectively (Fig.~\ref{fig:memory_traffic_breakdown}a).
However, HMS outperformed them by flexibly utilizing the bus across varying DRAM/SCM traffic ratios (Fig.~\ref{fig:memory_traffic_breakdown}b) and avoiding the external link bottleneck.

\noindent
{\bf \textit{Host interface impact.}}
With high-BW host memory access (in cacheline granularity for cold data), 
HBM(NVLink) outperformed InfHBM with PCIe by up to 96.4\% for workloads (e.g., \textfw{2DConv}, \textfw{pathfnd}) that did not thrash HBM (Fig.~\ref{fig:perf_75}).
However, when HBM was thrashed (e.g., \textfw{stencil}, \textfw{kcore}), it suffered from high page migration overhead.
Overall, HMS with PCIe outperformed HBM(NVLink) by 45\%. 
Since HMS is orthogonal to host interface choices, HMS(NVLink) was also evaluated and outperformed HBM(NVLink) by 2.11$\times$.

\noindent
{\bf \textit{BERT inference.}}
With HMS, GPUs can execute large language models that do not fit in HBM with high performance.
We evaluated inference of an enlarged BERT~\cite{bert} with 24.16~B parameters from 480 layers, which would fit in a GPU with 80~GiB HMS but not in the HBM of A100 40~GiB GPU. Thus, HBM GPU would fetch the model from the host with UM.
We evaluated its single middle encoder layer since all layers are identical except for the first and last layers, using TensorFlow XLA v2.4 and SQuAD~\cite{squad}. 
HMS outperformed HBM by 45.4\%, with only 1\% degradation than InfHBM. 
The DRAM cache hit rate of the HMS was 58\% overall and 96\% for writes, 
effectively reducing SCM writes. %

\begin{figure}
\minipage{0.46\linewidth}
\includegraphics[width=\linewidth]{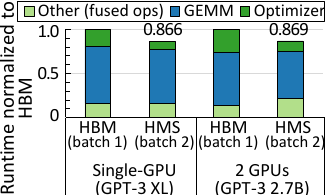}
\endminipage\hspace{0.1in}
\minipage{0.46\linewidth}
\includegraphics[width=\linewidth]{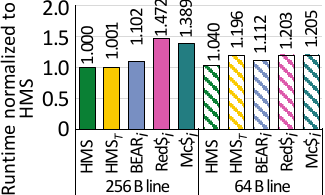}
\endminipage
\vspace{.1in}
\raggedleft{\hspace{1.7in}\footnotesize{(a)} \hspace{1.6in} \footnotesize{(b)}\hspace{0.6in}}
\caption{
(a) GPT model~\cite{gpt-3} training time with HBM and HMS using TensorFlow XLA.
(b) Performance impact of DRAM cacheline size (64 B vs. 256 B) with
different DRAM caches. HMS$_{T}$ is a variant of HMS with TAD instead of AMIL.}
\label{fig:cacheline_prior_works}
\end{figure}

\noindent
{\bf \textit{LLM training.}}
The high capacity of HMS can also benefit the training of LLMs such as GPT~\cite{gpt-3} on single or multi-GPU systems by enabling larger batch sizes, which reduces the optimizer runtime overhead and increases compute utilization~\cite{buddy_comp, megatron-lm}.
Due to prohibitively long runtime, a single decoder layer was simulated for comparison 
as all layers are identical.
Maximum possible batch sizes were used for each memory type, assuming 40 GiB HBM and 80 GiB HMS.
Single-GPU training used GPT-3 XL and 2-GPU training used GPT-3 2.7B with model parallelism~\cite{megatron}. 
For proper normalization of runtime, 2-iteration runtime with batch size of 1 for HBM was compared with single-iteration runtime with batch size of 2 for HMS. 
The HMS outperformed capacity-constrained HBM by 15.1\% (15.4\%) for 2-GPU (1-GPU) 
system (Fig.~\ref{fig:cacheline_prior_works}a).

\subsection{Sensitivity Study and Additional Results}
\label{sec:sensitivity_study}

\noindent
{\bf \textit{DRAM cacheline size impact.}}
Compared to using 64~B line, 256~B line provided 4\% speedups overall
for HMS due to improved CTC caching and the amortization of SCM latency (Fig.~\ref{fig:cacheline_prior_works}b).
HMS$_{T}$ benefited more from 256~B line (19.4\% speedup overall), as TAD
generates more DRAM traffic than AMIL from a CTC miss, and a larger line results 
in fewer tags.  
BEAR$_{i}$ improved little with 256~B line (0.9\% on average) while the performance of RedCache$_{i}$ and McCache$_{i}$ degraded with 256~B line, as they are unaware of SCM and increased SCM traffic further. 
In addition, for HMS, reducing cacheline size from 256~B to 128~B degraded performance by up to 11\% (1.7\% overall), while increasing it to 512~B had a negligible impact. 
1~KiB cacheline degraded performance by up to 6\% due to increased data movement.

\begin{figure}
\centering
\includegraphics[width=0.54\linewidth]{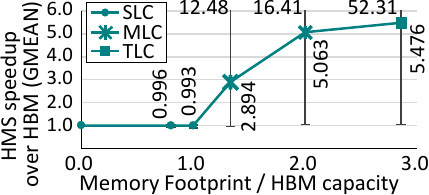}
\includegraphics[width=0.4\linewidth]{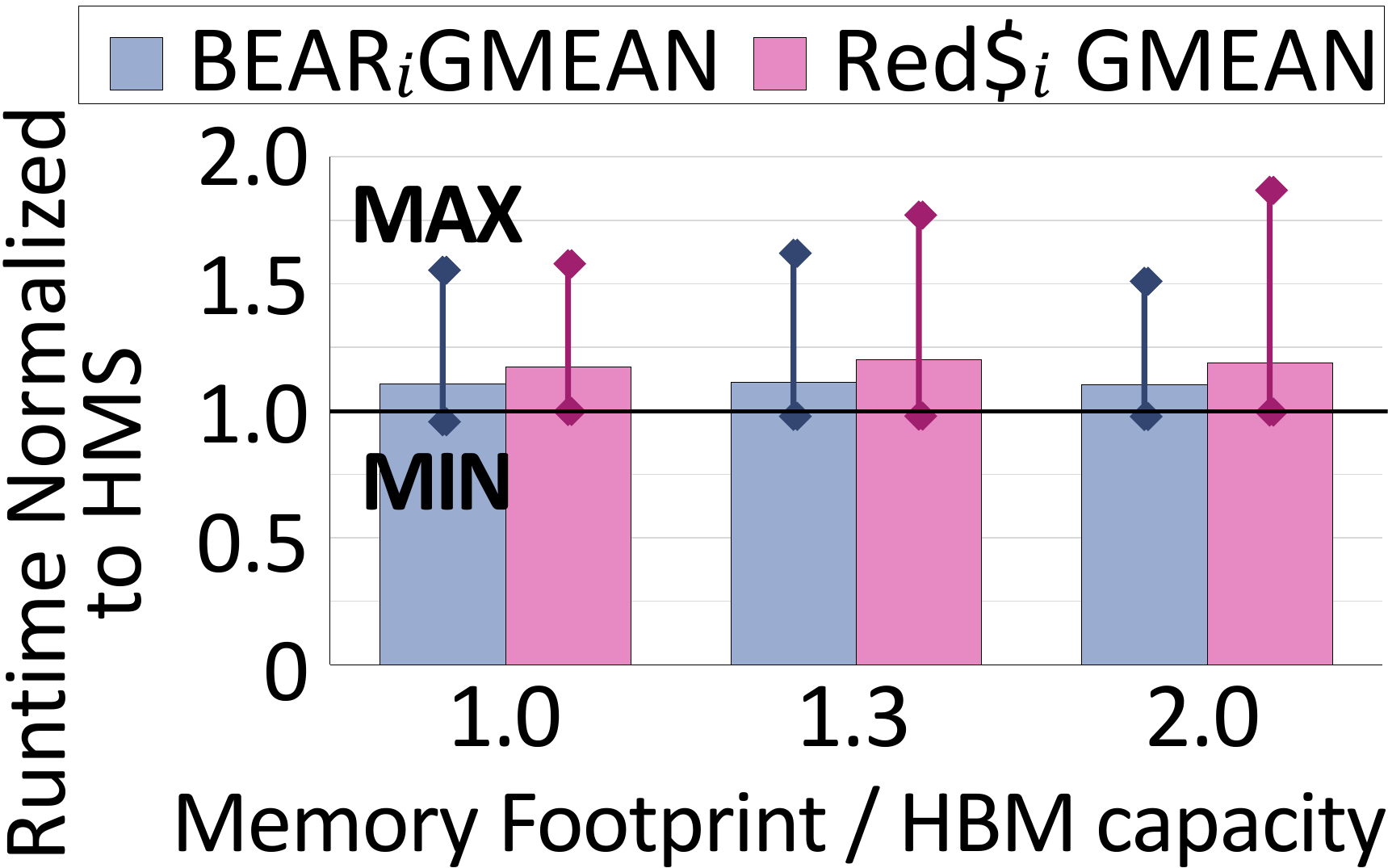}
\hbox{\hspace{0.5in}\footnotesize{(a)} \hspace{1.5in} \footnotesize{(b)}\hspace{0.1in}}
\caption{(a) Speedup with HMS over HBM for varying relative memory footprint over HBM capacity. The symbols in the legend indicate the mode of HMS's SCM among SLC, MLC, and TLC. Error bars show the maximum and minimum across workloads. (b) Runtime of BEAR$_{i}$ and RedCache$_{i}$ normalized to HMS for varying relative memory footprint for all workloads.}
\label{fig:workload_size}
\end{figure}

\noindent
{\bf \textit{Memory footprint impact.}}
Even for workloads with relatively small memory footprints, HMS showed competitive performance (within 1\%) compared to HBM by using DRAM as a part of memory and SCM in SLC mode (Fig.~\ref{fig:workload_size}a).\footnote{For SLC (TLC) SCM, we assumed RCD = 60 (250), RAS = 60 (250), and WR = 150 (2350)~cycles~\cite{scm_guideline}.}
As the relative memory footprint increases, HMS can use the SCM in TLC mode for higher capacity, achieving even greater speedups (up to 52.3$\times$ for \textfw{sssp\_dtc}).
Even when the HMS was oversubscribed with a relative footprint of 4.0, it still outperformed HBM by up to 108.8$\times$ (2.85$\times$ overall) by reducing page faults.
Thus, HMS can better serve diverse workloads than HBM. 
Additionally, for varied R$_{HBM}$, HMS also consistently outperformed Bear$_i$ (RedCache$_i$), by up to 62\% (87\%), except for \textfw{bckprp} for which Bear$_{i}$ outperformed HMS by only 1.2\%  (Fig.~\ref{fig:workload_size}b).

\begin{figure*} %
\minipage{0.324\textwidth}
\includegraphics[width=\linewidth]{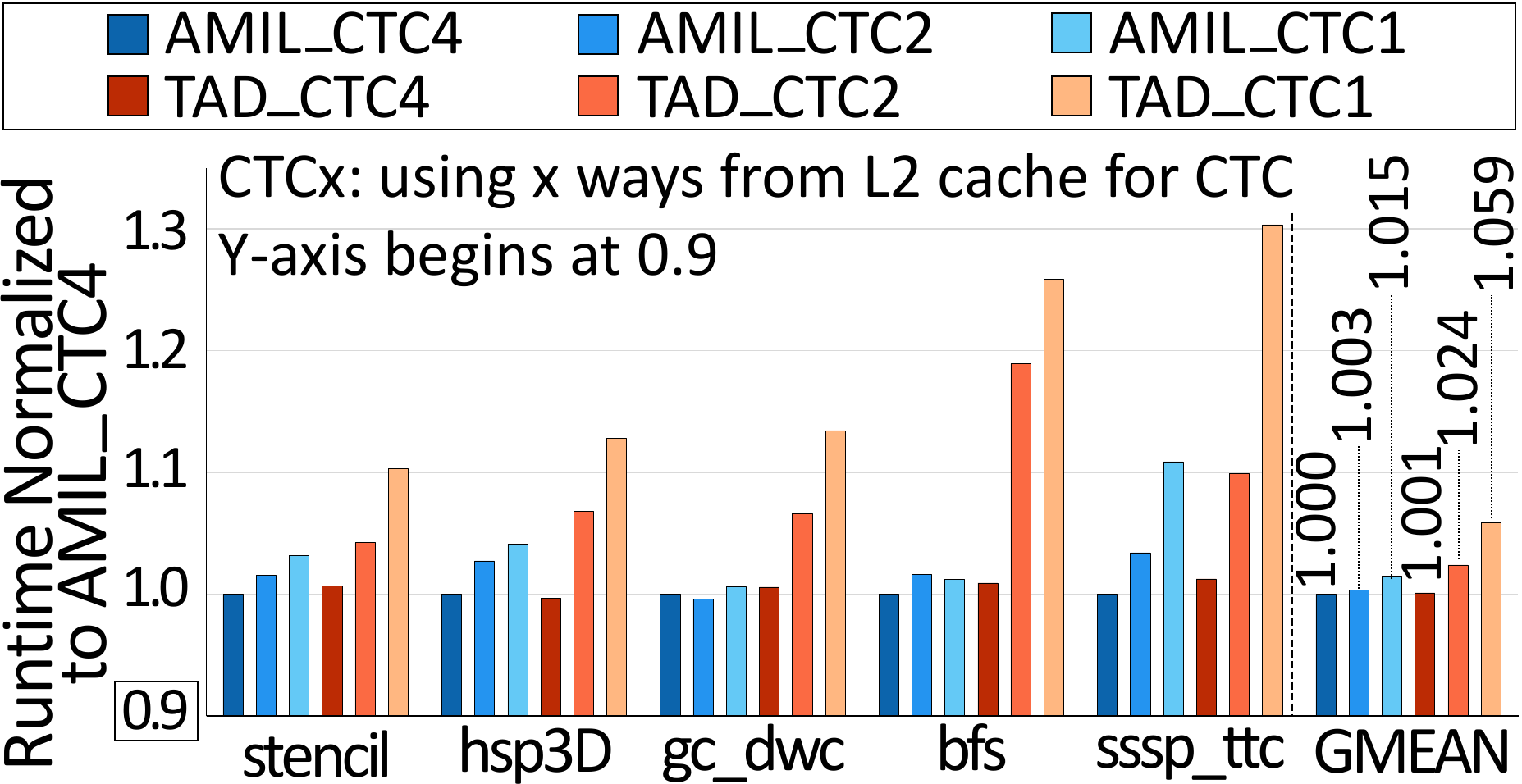}
  \caption{Performance impact of CTC ways.} 
  \label{fig:tag_cache_sensitivity}
\endminipage\hfill
\minipage{0.655\textwidth}
\includegraphics[width=\linewidth]{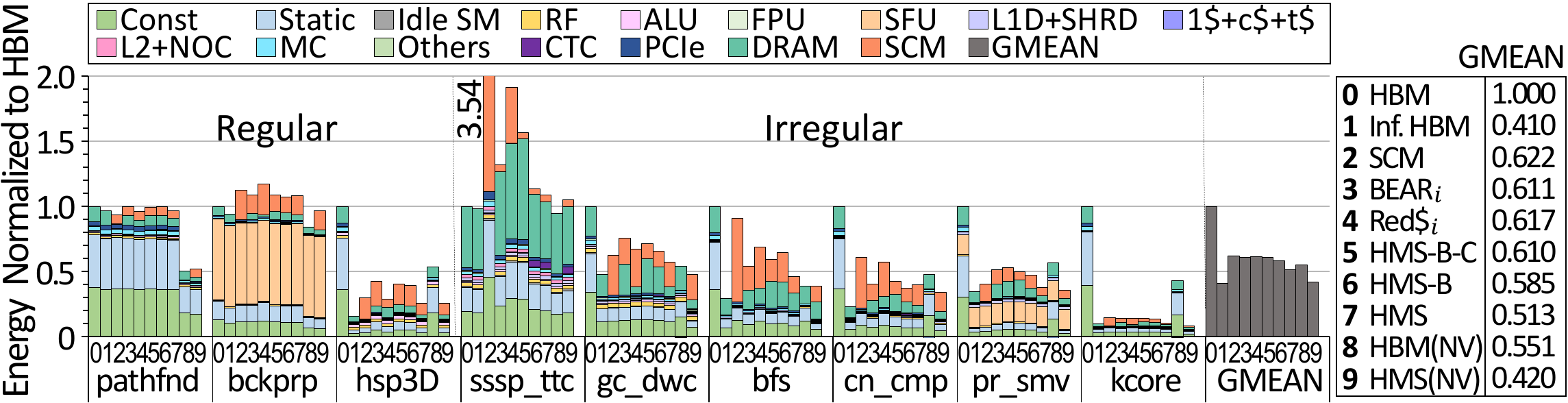}
  \caption{Energy consumption with different memory designs normalized to HBM.} %
  \label{fig:energy}
\endminipage\hfill
\end{figure*}

\noindent
{\bf \textit{CTC and AMIL sensitivity.}}
To analyze the impact of CTC size and AMIL, we varied the number of L2 cache ways used for CTC from 4 to 1 for AMIL and TAD (Fig.~\ref{fig:tag_cache_sensitivity}).
We assumed the same 256~B DRAM cacheline size to isolate its effect.
With AMIL, reducing the CTC size by a quarter had a low performance impact of only 1.5\% overall,
whereas TAD showed higher performance impact of 5.9\%.
AMIL outperforms TAD in handling the increased CTC miss, as AMIL needs a single DRAM access for a CTC miss, whereas TAD needs eight accesses due to distribution of tags in a row.
As a result, TAD\_CTC1 resulted in up to 5.6$\times$ (2.6$\times$ overall) more DRAM accesses than AMIL\_CTC1.

\noindent
{\bf \textit{DRAM vs. SCM capacity ratio impact.}}
We evaluated the effects of different capacity ratios between DRAM and SCM by varying their row counts.
For a configuration of 2 SCM dies and 6 DRAM dies (``2SCM-6DRAM''), runtime increased by up to 12.4$\times$ for \textfw{kcore} (2.9$\times$ overall), and energy increased by 1.99$\times$ compared to 4SCM-4DRAM due to smaller GPU memory capacity and frequent page faults. 
6SCM-2DRAM showed 6.5\% (10.4\%) higher runtime (energy) than 4SCM-4DRAM, as the DRAM cache hit rate decreased by 20.6\% overall due to smaller DRAM capacity.
For example, \textfw{hsp3D}'s DRAM cache hit rate fell by 49\% and the SCM activation increased by 35\%, resulting in a 29\% increase in runtime.

\noindent
{\bf \textit{PCIe BW and other sensitivities.}}
With 64~GB/s PCIe BW, HMS still outperformed HBM and SCM by 2.21$\times$ and 16.28\% overall, respectively.
HMS also still outperformed BEAR$_{i}$ (RedCache$_{i}$) by 16.13\% (19.89\%) overall.
Increasing $N_{levels}$ from 4 to 8 slightly degraded performance by 0.3\% overall due to increased traffic to probe  victim's DRAM affinity level.

\subsection{Energy and Power}

HMS substantially reduced energy consumption by up to 89.3\% (48.1\% overall) compared to HBM (Fig.~\ref{fig:energy}\footnote{In AccelWattch~\cite{accelwattch}, ``Static'' refers to energy from leakage currents of inactive components, and ``Const'' refers to peripheral component energy such as GPU board fans and other auxiliary circuitry.}) by reducing data movement and runtime.
Compared to SCM, HMS also considerably reduced energy by up to 68.0\% (16.5\% overall) as our DRAM cache effectively mitigated the high energy cost of SCM accesses. 
SCM-agnostic BEAR$_{i}$ and RedCache$_{i}$ did not measurably reduce energy, 
consuming 10.7\% and 76.8\% more energy on SCM access than HMS, respectively.
They also consumed 74.7\% and 7.4\% more energy on DRAM access overall than HMS due to frequent DRAM cacheline movements and tag probes. 
HMS(NVLink) also reduced the energy by up to 80.1\% (22.7\% overall) compared to HBM(NVLink).

\begin{figure}
    \centering
    \includegraphics[width=\linewidth]{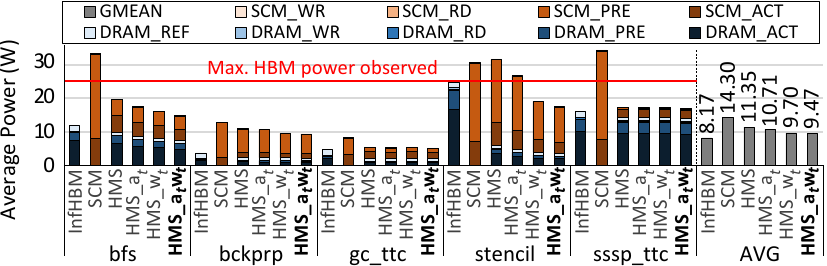}
    \caption{Average power usage by different memory stacks for representative workloads.
    For HMS, $a_t$ ($w_t$) indicates power throttling by doubling the corresponding timing parameters for activation (write recovery).} 
    \label{fig:power}
\end{figure}

While SCM can increase power usage, our simple SCM throttling technique (\S\ref{sec:opt}) can effectively prevent the power usage of HMS from exceeding the maximum power of HBM (Fig.~\ref{fig:power}).
In our evaluation, \textfw{stencil} showed the highest power consumption for InfHBM. 
While HMS without throttling consumed more power for \textfw{stencil},
using throttling effectively reduced the power to 54.5\% below that of InfHBM.
It resulted in 55\% performance loss but still outperformed the baseline HBM by 7.1$\times$.
In addition, even without throttling, HMS did not show high temperature 
(\S\ref{sec:thermal}) and other workloads did not require throttling. 
Thus, power and temperature of HMS can be safely managed.

\subsection{Thermal Model}
\label{sec:thermal}

\begin{table}
\caption{Thermal material properties.}
\label{tab:thermalconfig}
\scriptsize
\setlength{\tabcolsep}{2.0pt} %
\begin{center}
\vspace{-0.15in}
    \renewcommand{\arraystretch}{1.20}
    \begin{tabular}{|l|c|}
    \hline
    \multicolumn{2}{|c|}{\textbf{GPU}: Conductivity: 141W/m·K, Die size: 166mm$^2$, Power: 80W~\cite{a100}} \\ 
    \hline
    \hline
    \multicolumn{2}{|c|}{\textbf{Memory (a single stack with 8 memory dies and a base die)}} \\
    \hline
    \multicolumn{2}{|c|}{Conductivity(W/m·K): 141(base/DRAM~\cite{hotspotsim, DRAMthermal}), 106(SCM~\cite{PCMthermal}), 1.5(bonding)} \\
    \multicolumn{2}{|c|}{Base die power: 10W~\cite{basediepower1, basediepower2}, Die size: 96mm$^2$} \\
    \multicolumn{2}{|c|}{HBM: 8 DRAM dies, HMS: 4 DRAM (bottom) + 4 SCM (top) dies} \\
    \hline
    \hline
    \multicolumn{2}{|c|}{Convection resistance: 0.143K/W, Heat spreader: 400W/m·K, Ambient: 50°C} \\
    \hline
    \end{tabular}
\end{center}
\vspace{-0.1in}
\end{table}

As SCM can consume more energy than DRAM~\cite{lee2009architecting},
we evaluated the thermal behavior of different memory stacks using HotSpot thermal modeling tool~\cite{hotspotsim}
(Table~\ref{tab:thermalconfig}).
The thermal model includes a silicon interposer, GPU die, base die, memory dies, bonding layers (between memory dies), and cooling solution with a general heat spreader and air-cooling heat sink.
We conservatively assumed that the GPU consumes the TDP (i.e., maximum sustainable power) of the scaled-down NVIDIA A100 and that the base die of HBM consumes the TDP of 10W~\cite{basediepower1, basediepower2}.

\begin{figure}
\centering 
\includegraphics[width=1.0\linewidth]{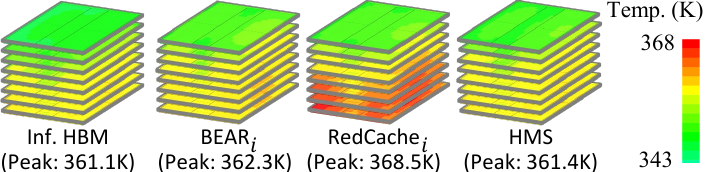}
\caption{Thermal maps for \textfw{stencil} (worst-case thermal behavior among evaluated workloads). The GPU, base die, and bonding layers of the stacks are included in the thermal model but omitted in the figure for brevity.}
\label{fig:thermal}
\end{figure}

\begin{figure}
    \centering
    \includegraphics[width=1.0\linewidth]{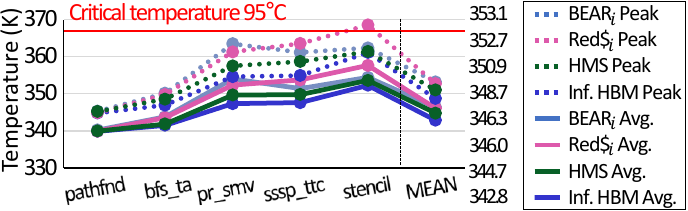}
    \vspace{-0.1in}
    \caption{Peak and average temperatures of different DRAM caches and InfHBM.} %
    \label{fig:thermal_graph}
\end{figure}

For \textfw{stencil}, which showed the highest power usage, HMS showed similar thermal behavior to InfHBM, while RedCache$_{i}$ exceeded the 95°C critical temperature due to high DRAM traffic (Fig.~\ref{fig:thermal}).
In 3D memory, the bottom die has the poorest heat dissipation as it is farthest from the heat sink.
Despite consuming more power in the SCM dies than the DRAM dies in HBM, HMS had a lower DRAM power usage that resulted in a negligible increase in peak temperature.
In addition, HMS resulted in lower average and peak temperatures than prior works for all workloads in our evaluation (Fig.~\ref{fig:thermal_graph}).

\subsection{Hardware Overhead}
\label{sec:hardware_overhead}

HMS requires a 128-entry MSHR per channel, and each entry requires 51 bits (37-bit address, 8-bit mask to record columns accessed in a cacheline, entry valid bit, read/write bit, 2-bit DRAM affinity level, and the DRAM cacheline valid and dirty bits from CTC). 
Using CACTI~\cite{cacti7} and assuming 12nm, the MSHR and
256-bit storage for our bypass logic is estimated to use 0.0006~mm$^2$ per memory channel. 
The overhead of CTC per memory partition, including comparators and muxes, is estimated to be 0.014~mm$^2$.
An integer ALU (\S \ref{sec:dev_sensitivity}) and an FPU per channel have 
an area of 0.022~mm$^2$ in 12~nm~\cite{FPnew}.
Overall, the overhead with 40 memory channels in an NVIDIA A100 GPU is estimated to be 1.46~mm$^2$ or a 0.18\% increase. 
This area estimation with 12nm is conservative 
given that the A100 GPU used 7nm technology~\cite{a100}.

\section{Related Work}
\label{sec:related_work}

\subsection{DRAM Cache}
Many DRAM cache designs have been proposed, especially for CPUs with both high- and low-BW DRAM. 
Alloy Cache~\cite{alloycache} proposed direct-mapped DRAM cache with TAD organization that trades off hit rate against hit latency in comparison to Loh-Hill cache~\cite{loh_hill}. 
Timber~\cite{timber} and ATCache~\cite{atcache} proposed a fixed-size on-chip SRAM storage for DRAM cache tags while our CTC allows size configuration by users.
ACCORD~\cite{accord} mitigated high BW overhead of set-associative DRAM caches by coordinating way install and prediction.
Tag Tables~\cite{tagtables} modifies page table to compress DRAM cache tags and 
cache them in LLC. 
Footprint-based DRAM caches~\cite{footprint_cache, ftdc, unisoncache} 
exploits \emph{intra-thread} spatial locality of CPU threads (\S \ref{sec:considerations}).
BEAR~\cite{bear} addressed DRAM cache's BW bloat with probabilistic bypassing, write probe filtering with metadata in LLC, and fetching/caching neighbor DRAM cacheline tags for demand accesses. 
RedCache~\cite{redcache} bypasses DRAM cache with dynamic access count thresholds to identify hot data. 
DICE~\cite{dice} is a dynamic cacheline indexing scheme for compressing DRAM caches.
Baryon~\cite{baryon} uses compression and sub-blocking to efficiently utilize fast memory capacity with low BW overhead.
These compression schemes can be adopted in our DRAM cache to further improve effective BW. 
Several works~\cite{carve, gpudmm, c3d} proposed DRAM cache for remote data
in other GPUs or CPUs. 
PoM~\cite{pom} and CAMEO~\cite{cameo} proposed using stacked DRAM to expand address space, rather than as a cache. 
Page-granularity DRAM cache management has also been proposed~\cite{banshee, tdc, hetero_mem}.
Sim et al.~\cite{mostly_clean} proposed keeping DRAM cache mostly clean.
However, they did not consider SCM's characteristics.

\subsection{Hybrid and Adaptive Memory Hierarchy}
Several prior works~\cite{pcm_isca09, pram_3dstack, rowlocality} proposed memory systems with DRAM and PCM, optimizing for page management and endurance.
Yoon et al.~\cite{rowlocality} and Zhao et al.~\cite{zhao2012optimizing} proposed data placement mechanisms considering row buffer miss frequency similar to our hotness metric, but they did not consider the inter-thread spatial locality of GPUs.
They also managed the DRAM cache in a large row-granularity, but smaller DRAM cachelines are more effective for GPUs (\S\ref{sec:sensitivity_study}). 
3D-Xpath~\cite{3dxpath} proposed a 3D memory stack combining density-optimized and performance-optimized DRAM.
Memory hierarchies using sub-ranks~\cite{subrank} or sub-channels~\cite{subchannel} can improve energy efficiency by finer granularity accesses. These approaches are orthogonal to our design and can be combined.
Ohm-GPU~\cite{ohm-gpu} proposed a silicon photonics-based optical network 
for GPU, DRAM, and 3D XPoint memory for high BW. 
However, our work focuses on a more practical near-term solution.
ZnG~\cite{ZnG} and FlashGPU~\cite{flashgpu} integrated flash devices in the memory hierarchy, which can be effective for read-intensive, regular access patterns.
However, for frequent irregular writes, the high ($\sim$100$\mu$s) write latency and granularity would require an effective DRAM cache. Our DRAM cache design can be adopted in their designs to improve performance.
Thermostat~\cite{thermostat} is a SW scheme to manage data placement between SCM and DRAM under user-specified performance constraints.
Kim et al.~\cite{nomad} and Liu et al.~\cite{liu2019hierarchical} also proposed hybrid memory systems managed by the OS.
However, for GPU workloads, SW-managed DRAM caches can become a bottleneck. 
Several works~\cite {optane_eurosys22, optane_ispass20, optane_micro20} analyzed Optane PM DIMM and proposed SW optimizations to improve performance. Optane DIMM does not necessarily represent the SCM we assume, as it includes a hierarchy of buffers that access 3D XPoint memory with a large 4~KiB granularity. Moreover, requests can be reordered by an on-DIMM queue.
Recent key-value stores~\cite{kv1,kv2,eisenman2018reducing} exploited heterogeneous memory hierarchy.
Harmony~\cite{harmony} proposed scheduling of tasks and data movement for
training large DNNs on a GPU. 
GPM~\cite{gpm} exploits the persistency of CPU-attached Optane from GPU.
MMS~\cite{morphable_isca10} proposed HW/OS support for adapting 
between high-density and low-latency PCM modes.
Power token~\cite{pcm_power} was proposed to manage PCM power 
with fine-grained write. %

\section{Conclusion}

We propose an effective DRAM cache for GPUs with SCM to overcome the
memory capacity wall while achieving high memory BW.
Our AMIL organization fetches
all tags in a row with a single access to reduce the tag probe BW overhead, while 
retaining full 
ECC protection in contrast to prior DRAM caches with TAD organization.
Furthermore, to prevent DRAM cache thrashing from a massive number of threads while considering the characteristics of SCM, we propose an \emph{SCM-aware DRAM cache bypass policy}.
This policy leverages the SCM penalty score and DRAM-affinity score, which captures the multidimensional characteristics of access patterns (i.e., access frequency, row buffer locality, and write intensity) in a single score, for simple yet effective bypassing.
In addition, because DRAM cache probe traffic can interfere with data access from DRAM, we propose CTC to reduce the probes with little overhead while enabling flexible capacity adjustment between CTC and L2 cache. 
Consequently, we reduce DRAM cache probe and SCM write traffic by 91-93\% and
57-75\%, respectively, over prior works.
Our SCM throttling can effectively curtail SCM power usage below 
the maximum HBM power
while still achieving high speedups over HBM.
Using SCM's SLC and MLC modes, the GPU can also adapt to workload's memory footprint
and performance demand.
The results show that our proposed GPU with SCM and DRAM cache 
significantly outperforms the oversubscribed baseline GPU with HBM 
by up to 12.5$\times$ (2.9$\times$ overall).

\section*{ACKNOWLEDGMENT}
This work was supported by Institute of Information \& communications Technology Planning \& Evaluation (IITP) grants
(No.2021-0-00871, Development of DRAM-Processing-In-Memory Chip for DNN Computing, 
and No.2021-0-00310, Development of SW Framework for Server to Improve AI Training/Inference Efficiency)
and National Research Foundation of Korea (NRF) grants
(RS-2023-00277080 and RS-2023-00212711) funded by the Korea government (MSIT).
We would like to thank the anonymous reviewers for their constructive comments.
Gwangsun Kim is the corresponding author.

\balance
\bibliographystyle{IEEEtranS}


\end{document}